\documentstyle[emulateapj,pstricks]{article}

\def\Msun{~M_{\odot}\ }
\def\kpc{\rm kpc}
\def\Mhsun{h^{-1}{\ }{\rm M_{\odot}}\ }
\def\Mpc{{\rm Mpc}}
\def\Mpch{h^{-1}{\rm Mpc}}
\def\hMpc{h^{-1}{\rm Mpc}}
\def\kpch{h^{-1}{\rm kpc}}
\def\hkpc{h^{-1}{\rm kpc}}

\def\kms{~\hbox{km\ s}^{-1}}
\def\M10{{\times 10^{10} M_{\odot}\ }}

\def\Mvir{M_{\rm vir}}
\def\Rvir{r_{\rm vir}}

\def\SecDynFric{2.4}

\def\SecBDM{4.2}

\def\EqNFW{1}
\def\EqVmax{5}
\def\EqTidal{6,7}
\def\EqDynFr{8} 
\def\Eqtfric{8} 
\def\EqVelDisp{10}
\def\Tideone{1}
\def\TideIso{2}
\def\TideResolution{3}
\def\TideOrbit{4}
\def\TideOrbits{5}
\def\DVevol{6}
\def\Dynfriction{7}
\def\HalosGr{8}
\def\Groupz{9}
\def\Halos{10}
\def\Profsm{11}
\def\Profmed{12}
\def\LumLCDM{13}
\def\LumCHDM{14}
\def\MassVelocityLCDM{15} 
\def\MassVelocityCHDM{16}
\def\CorrFun{17}
\def\CorrMass{18}

\def\mathnew{\mathsurround=0pt}
\def\ref{\par\noindent\hangindent=2pc \hangafter=1 }
\def\simov#1#2{\lower .5pt\vbox{\baselineskip0pt
    \lineskip-.5pt\ialign{$\mathnew#1\hfil##\hfil$\crcr#2\crcr\sim\crcr}}}

\def\'#1{\ifx#1i{\accent"13\i}\else{\accent"13#1}\fi}

\begin{document}
\slugcomment{{\em Astrophysical Journal, in press}}
\lefthead{}
\righthead{KLYPIN ET AL.}

\title{GALAXIES IN $N$-BODY SIMULATIONS:\\
OVERCOMING THE OVERMERGING PROBLEM}\vspace{3mm}

\author{Anatoly A. Klypin, Stefan Gottl\"ober\altaffilmark{1}, and Andrey V. Kravtsov}
\affil{Astronomy Department, New Mexico State University, Box 30001, Dept.
4500, Las Cruces, NM 88003-0001}
\altaffiltext{1}{On leave from Astrophysikalishes Institut Potsdam
  (AIP), An der Sternwarte, 16, D-14482, Potsdam, Germany}
\vspace{1mm}
\author{Alexei M. Khokhlov}\vspace{2mm}

\affil{Laboratory for Computational Physics and Fluid Dynamics, Code
  6404, 
Naval Research Laboratory, Washington, DC 20375}

\begin{abstract}
  We present analysis of the evolution of dark matter halos in dense
  environments of groups and clusters in dissipationless cosmological
  simulations. The premature destruction of halos in such environments,
  known as the {\it overmerging}, reduces the predictive power of
  $N$-body simulations and makes difficult any comparison between
  models and observations. We analyze the possible processes that cause
  the overmerging and assess the extent to which this problem can be
  cured with current computer resources and codes.  Using both analytic
  estimates and high resolution numerical simulations, we argue that
  the overmerging is mainly due to the lack of numerical resolution.
  We find that the force and mass resolution required for a simulated
  halo to survive in galaxy groups and clusters is extremely high and
  was almost never reached before: $\sim 1-3$ kpc and $10^8-10^9\Msun$,
  respectively. We use the high-resolution Adaptive Refinement Tree
  (ART) $N$-body code to run cosmological simulations with the particle
  mass of $\approx 2\times 10^8h^{-1}{\ }{\rm M_{\odot}}$ and the
  spatial resolution of $\approx 1-2h^{-1}{\ }{\rm kpc}$, and show that
  in these simulations the halos do survive in regions that would
  appear overmerged with lower force resolution. Nevertheless, the halo
  identification in very dense environments remains a challenge even
  with the resolution this high.  We present two new halo finding
  algorithms developed to identify both isolated and satellite halos
  that are stable (existed at previous moments) and gravitationally
  bound.
  
  To illustrate the use of the satellite halos that survive the
  overmerging, we present a series of halo statistics, that can be
  compared with those of observed galaxies. Particularly, we find that,
  on average, halos in groups have the same velocity dispersion as the
  dark matter particles, i.e. do not exhibit significant velocity bias.
  The small-scale (100~kpc -- 1~Mpc) halo correlation function in both
  models is well described by the power law $\xi\propto r^{-1.7}$ and
  is in good agreement with observations. It is slightly antibiased
  ($b\approx 0.7-0.9$) relative to the dark matter. To test other
  galaxy statistics, we use the maximum of halo rotation velocity and
  the Tully-Fisher relation to assign the luminosity to the halos. For
  two cosmological models, a flat model with the cosmological constant
  and $\Omega_0=1-\Omega_{\Lambda}=0.3, h=0.7$, and a model with a
  mixture of cold and hot dark matter and $\Omega_0=1.0,
  \Omega_{\nu}=0.2, h=0.5$, we construct the luminosity functions and
  evaluate mass-to-light ratios in groups.  Both models produce
  luminosity functions and the mass-to-light ratios ($\sim 200-400$)
  that are in a reasonable agreement with observations.  The latter
  implies that the mass-to-light ratio in galaxy groups (at least for
  $M_{vir}\lesssim 3\times 10^{13}h^{-1}{\ }{\rm M_{\odot}}$ analyzed
  here) is not a good indicator of $\Omega_0$.

\end{abstract}
\keywords{cosmology: theory -- large-scale structure of universe --
  methods: numerical}


\section{Introduction}


It is generally believed that the dark matter (DM) constitutes a large
fraction of the mass in the Universe and thus significantly affects,
and on some scales dominates, the process of galaxy formation.
Observational evidence for large fractions of DM in galaxies, groups,
and galaxy clusters ranges from flat rotational curves of spiral (e.g.,
Faber \& Gallagher, 1979; Rubin et al. 1985; Persic, Salucci, \& Stel
1996; Courteau \& Rix, 1997) and X-ray emission and mass-to-light
ratios of elliptical (Forman et al.1985; Rix 1997; Brighenti \& Mathews
1997) galaxies to the baryon fractions in clusters of galaxies (White
et al. 1993; Evrard 1997). For
galaxies, the extent of the DM halos estimated using satellite
dynamics, is $\sim 0.2-0.5h^{-1}$ Mpc\footnote{Throughout this paper we
  assume that the present-day Hubble constant is $H_0=100h{\ }{\rm km
    s^{-1} Mpc^{-1}}$.} (Zaritsky \& White 1994; Carignan et al.1997).
A convincing evidence for substantial amounts of dark matter
even in the very inner regions of galaxies comes from the recent HI
studies of the dwarf and low surface brightness (LSB) galaxies.  The
observed amounts of stars and gas in some of these galaxies can account
for less than $10\%$ of the observed rotational velocities at the last
measured point of the rotation curves.  (e.g., Carignan \& Freeman
1988; Martimbeau, Carignan, \& Roy 1994; de Blok \& McGaugh 1997).

If the observed galaxies  have large DM halos, then $N$-body
simulations can, in principle, be used to predict distribution of the
dark matter component, to associate the simulated DM halos with
galaxies, and to predict the bulk properties of these galaxies such as
position, mass, and size.  One should be able then to make predictions
about spatial distribution and motion of these simulated galaxies and
compare these predictions with corresponding observations.
Unfortunately, the dissipationless numerical simulations have been
consistently failing to produce galaxy-size dark matter halos in dense
environments typical for galaxy groups and clusters (e.g., White 1976;
van Kampen 1995; Summers, Davis, \& Evrard 1995; Moore, Katz, \& Lake
1996).  This apparent absence of substructure in the virialized
objects, known as {\it the overmerging problem}, reflects the fact that
simulated galaxies seem to merge much more efficiently in comparison
with real galaxies in groups and clusters. In the central regions of a
cluster ($\sim 500$ kpc ), the ``overmerging'' erases not only
large-scale substructure, but also any trace of small halos that could
be associated with ``galaxies''\footnote{The term ``galaxy''
  traditionally refers to luminous observed objects (i.e. to clumps of
  stars and gas, not the DM), which, possibly, are embedded in a
  considerably larger DM ``halo''. The term ``halo'', however, is
  rather general. We can use this term to indicate a galaxy cluster,
  group, or a galaxy-size halo. In some cases, we want to make a clear
  distinction between these. We will thus use terms ``simulated
  galaxies'' or ``galaxy-size halo'' to indicate the DM halos formed in
  the simulations which could be associated with places where luminous 
  baryons could reside.}, leaving a smooth giant lump of dark
matter.

The overmerging problem was traditionally explained by the lack of
dissipation in $N$-body simulations (e.g., Katz, Hernquist \& Weinberg
1992; Summers et al.  1995). Indeed,
the DM halos are much larger than baryonic extent of the galaxies due
to the dissipational nature of the latter. The radiative cooling, for
example, allows baryonic component to sink into the center of the DM
halo where it forms a compact, tightly bound object.  In dense
environments the large DM halo can be easily stripped by the
tidal field of a galaxy cluster or group, whereas the more compact and
denser gas clump may survive (Summers et al.  1995).  Although it is
clear that to produce a realistic galaxy we need to include the energy
dissipation by baryons, it is not clear whether the dissipation is vital for
the halo survival in a cluster.  

Two arguments can be presented against the traditional explanation for
the overmerging.  First, if the dissipation helps galaxies to survive
in clusters, then galaxies should be dominated by baryons at all scales
within their visible extent.  Most of the observed galaxies, however,
appear to have a substantial fraction of DM inside their optical radius
(e.g., Persic et al.  1996).  The survival of a galaxy dominated at its
optical radius by the dark matter will depend mostly on the dark
matter, not on the baryons. The DM dominated dwarfs must have been
tidally disrupted, but dwarf galaxies are observed in
clusters (e.g. Smith, Driver, \& Phillipps 1997; Lopez-Cruz et al. 
1997). Second, as we argue in this paper, even in the absence of the
baryons DM halos are dense enough to survive inside clusters and to be
identified, provided that simulations have sufficient resolution {\em
  in both} mass and force (similar conclusion was reached by Moore et
al. 1996).

The main goal of this paper is to demonstrate that with a sufficient
computational effort the overmerging problem can be tamed (at least to some
extent) even in the purely dissipationless simulations. The
computational costs are higher than cost of an average cosmological
$N$-body simulation. However, they may be  considerably lower than
computational expense of the corresponding $N$-body$+$hydro
simulations. This especially true in the case of large-volume ($\sim
50-100h^{-1}{\ }{\rm Mpc}$) simulations required to quantify the statistical
properties of the ``galactic'' populations such as correlation
function, pairwise velocity dispersions etc. To understand how the
problem should be dealt with, it is important to understand what
processes lead to the overmerging.

Several recent studies have addressed this question from different
viewpoints and using different numerical and analytical techniques.
Thus, for example, van Kampen (1995) studied formation of galaxy
clusters in purely dissipationless simulations and concluded that
two-body relaxation and tidal disruption are primarily responsible for 
the overmerging. He found, however, that the two-body effects are
important only for the smallest halos ($\lesssim 30$ DM particles), in
quantitative agreement with experiments of Moore et al. (1996). The
latter study addressed also effects of particle-halo and halo-halo
heating on the survival of DM halos. The authors concluded that
particle-halo heating does not pose a problem as long as DM particle
mass is $\lesssim 10^{10} M_{\odot}$, but halo-halo heating may be
important if force resolution is not adequate ($\gtrsim 10 {\rm kpc}$). The
general conclusion of Moore et al. is that the overmerging problem is
due mainly to tidal heating by the cluster and halo-halo heating, both
effects being enhanced by poor force resolution. They note also
that these effects depend crucially on the density structure of DM
halos.

The density profile of dark matter halos in the CDM models is now known
reasonably well.  Navarro, Frenk \& White (1995, 1996, 1997, hereafter
NFW) gave an analytical fit that describes with a reasonable accuracy
the density profiles of DM halos formed in the standard cold dark
matter scenario over large range of scales and masses. The analytical
form of the density profile advocated by NFW, allows one to estimate
tidal effects and effects of dynamical friction analytically. We
present these estimates in \S 2.

Two recent numerical studies (Tormen, Diaferio \& Syer 1998; and Ghigna et
al. 1998) present evidence that higher resolution significantly
aleviates the overmerging problem in dissipationless simulations. Both
studies conclude that many halos survive for a long time after falling
onto a large halo, although overmerging problem persists to a certain
degree in the central dense region of the large cluster-size halo. 
Tormen et al. also point out that there are real physical processes
which would lead to the erase of substructure even in the ideal very
high-resolution simulation. Namely, the dynamical friction drives
massive satellites to the cluster core where they get tidally stripped
and quickly disrupted. This effect is, of course, real and should be 
distinguished from artificial overmerging caused by insufficient
numerical resolution. We will address these issues in \S 2. 
 
Two goals of the study presented in this paper are 1) to make an
approximate estimates of the effects leading to the overmerging for the
halos with the NFW density distribution; and 2) to demonstrate that
dissipationless simulations with sufficiently high resolution in force
and mass are affected by the overmerging to a considerably lesser
degree.  In other words, we present an attempt to estimate to which
extent the overmerging problem can be solved with numerical resolution
that can be achieved with current codes and computational resources.
The goal is worthwhile. If the problem can be minimized at a
computational cost that is not prohibitive, the dissipationless
simulations can be used for a direct study of the statistical
properties of ``galaxies''. This may include studies of their spatial
distribution, velocity field, environmental effects etc.  This may
allow us to make a step towards solution of the long-standing and
particularly important issue of the galactic bias.

The overall plan of the paper is as follows.  In \S 2 we discuss the
numerical and physical effects which may lead to the erasure of
substructure inside the dense massive halos. Specifically, we present
analytical estimates for tidal disruption of dark matter halos and
effects of dynamical friction assuming that halos are described by the
NFW density profile. We use these calculations to make a rough estimate
of what resolution would be required to minimize the overmerging.
Numerical simulations and cosmological models are discussed in \S 3. In
\S 4 we discuss difficulties associated with identification of dark
matter halos in very dense regions. We describe two new halo finding
algorithms developed to handle the halos in such environments. In \S 5
we present results of high-resolution dissipationless simulations,
illustrate the performance of the new halo finding algorithms and
discuss the degree to which the simulations are affected by the
overmerging. Our conclusions are illustrated using well-known
statistics such as two-point correlation function, velocity bias,
luminosity function and $(M/L)$ ratio. The conclusions are summarized
in \S 6.

\section{SURVIVAL OF HALOS IN CLUSTERS}

\subsection{Numerical effects vs. physical effects}

As was noted in the introduction, there is a number of processes 
potentially contributing to the erasure of substructure in clusters and
groups (van Kampen 1995; Moore et al.  1996). Some of these processes are
due to numerical limitations of a simulation, while others are real
physical effects. 

The major effects caused by numerical limitations are {\em particle
  evaporation} due to the two-body relaxation (e.g., Carlberg 1994; van
Kampen 1995; Moore et al. 1996), {\em particle-halo heating} (Carlberg
1994; Moore et al. 1996), and {\em premature tidal disruption} due to
an insufficient force resolution. The two-body evaporation in an
$N$-body system is a well-understood process (e.g., Binney \& Tremaine
1987).  However, it was shown (van Kampen 1995; Moore et al. 1996) that
the two-body evaporation is important only for the halos with $\lesssim
30$ particles. Thus, if a simulation has a particle mass of $\lesssim
10^9h^{-1}{\ }{\rm M_{\odot}}$, this effect is not important for halos
in the mass range of interest ($\gtrsim 10^{11}h^{-1}{\ }{\rm
  M_{\odot}}$). It may become important, however, if halo looses most
of its mass due to the tidal stripping. The particle-halo heating was
suggested by Carlberg (1994) as an explanation for overmerging in his
simulation. However, Moore et al. (1996) showed that the time-scale for
this process is large for typical numerical parameters. Finally,
gravitational softening, $r_{soft}$, imposed in an $N$-body simulation
usually leads to a constant density core $r_c\approx r_{soft}$ in the
halo center. Poor force resolution may, thus, result in halos with
artificially large cores which will be tidally disrupted faster than if
they would have a more compact density distribution.

The real physical processes which may lead to the erasure of
substructure include {\em dynamical friction}, {\em tidal stripping},
and {\em halo-halo heating}. The dynamical friction drives DM halos
towards the high-density cluster center where they get tidally stripped
and merge with the central massive object. Dynamical friction can be
important for some halos and is probably responsible for the presence
of massive central cD galaxies in observed clusters (e.g., Merritt
1985). The importance of dynamical friction for each particular halo
will depend on the halo mass and details of its orbit. We will estimate
these dependencies in \S {\SecDynFric}. Tidal stripping occurs simply
because at some distance from the center the tidal force from a cluster
is stronger than gravitational force of a parent halo and particles
beyond this radius become unbound from a halo and dissolve in the
ambient diffuse medium.  This effect will be discussed in the next
section. Finally, Moore et al. (1996) argued that heating due to the
close passages of DM halos on their orbits may lead to significant
mass losses. Their estimate, however, was based on experiment in which
cluster is static.  In real clusters only a few halos will be
present for its entire lifetime, the bulk of the halos being
accreted over period of the cluster evolution. This process is probably less
important than the more efficient tidal stripping. 

Numerical and physical effects are, of course, closely related.  Tidal
force of the cluster and close encounters of individual dark halos
result in effective stripping of the peripheral parts of the halos.
This is a real physical effect. But after the stripping is done,
two-body relaxation can become artificially short and can result in
evaporation of halos if the number of particles in a halo is too small.
Due to the dynamical friction halos tend to spiral down towards the
center of the cluster where they merge with the central halo. This is a
real effect. But it can be exacerbated by artificially high energy
dissipation in hydrodynamical simulations resulting in too compact
halos.     

It is clear that physical processes will operate and will tend to erase
substructure even in an ideal simulations of infinite resolution.  The
numerical effects, on the other hand, can be cured by improving the
spatial and mass resolution of the simulations and/or by inclusion of
additional physics (e.g., missing gas dissipation). It is the numerical
effects that we will focus on in this paper. Throughout the rest of the
paper, we will be using the term ``overmerging problem'' only for
processes related to the numerical effects, which appear due to the
lack of resolution or due to missing physical effects.

\subsection{Tidal disruption of halos: analytical estimates}

The main reason for the erasure of the substructure in clusters is the
tidal interaction of individual halos with the cluster potential. This
can be now estimated reliably, without assuming that halos are
isothermal spheres or have King profiles, as was typically the case in
the past (Moore et al. 1996). The density profiles of dark matter halos
in the CDM models is now known reasonably well for a large range of
masses and for a variety of cosmological models (NFW). Below we give
analytic estimates for various halo properties in the standard $\Omega=1$ CDM
model. The low-density flat CDM model with cosmological 
constant ($\Lambda$CDM; e.g., Carrol, Press \& Turner 1992) model predicts
similar profiles, if the the overdensity of a collapsed object is
adjusted properly to take into account the change in $\Omega$ (Lahav
et al. 1991; Eke et al. 1996).

The NFW density profile is given by
\begin{eqnarray}
\rho(r) & = & {\rho_0 r_s^3\over r(r+r_s)^2}, \quad
M(r)  =  \Mvir\cdot{f(x)\over f(C)}, \\
f(x) &\equiv &\ln(1+x) -{x\over 1+x}, \quad x \equiv {r\over
r_s}. \nonumber
\end{eqnarray}
where $r_s$ and $\rho_0$ are the characteristic radius and density of the halo,
$\Mvir$ is the virial mass, $\Rvir$ is the virial radius, and
$C$ is the concentration for a halo defined as follows:
\begin{eqnarray}
C & \equiv &{\Rvir\over r_s}, \nonumber \\
\Rvir(\Mvir) & = & 443\hkpc\left(\Mvir/10^{11}\Mhsun\over
            \Omega_0\delta_{\rm th}\right)^{1/3}, \\ 
\Mvir  & \equiv &
{4\pi\over 3}\rho_{cr}\Omega_0\delta_{\rm th}\Rvir^3.
                             \nonumber   
\end{eqnarray}
 Here, $\rho_{cr}$ is the critical density of the Universe and
$\delta_{\rm th}$ is the overdensity ($\delta\rho/\rho_{\rm matter}$)
of a collapsed object according to the top-hat model of spherical
collapse. For the CDM model $\delta_{\rm th}\approx 200$. Note that
our definition of $\Rvir$ differs from that used by NFW,
 who use $\delta_{\rm th}=200$ for all cosmological
models. We use values predicted by the top-hat model which gives, for
example, $\delta_{\rm th}\approx 340$ for the $\Lambda$CDM model with
$\Omega_0=0.3$.

The concentration $C$ is a function of mass $\Mvir$. In the case of the
CDM model 
\begin{equation}
  C\approx 124(\Mvir/1\Mhsun)^{-0.084}
\end{equation}
Typical values for the
concentration $C$ range from $C\approx12$ for $\Mvir=10^{12}\Mhsun$ to
$C\approx 7$ for $\Mvir=10^{15}\Mhsun$ (NFW).
Using these definitions we can
write mass $M(r)$, orbital frequency $\omega(r)$, and gravitational
potential $\phi(r)$:
\begin{eqnarray}
\omega^2(r) & = & {GM\over r^3} = {G\Mvir\over r_s^3f(C)}\cdot
            {f(x)\over x^3},\nonumber\\
\phi(r) &= &-{G\Mvir\over r_sf(C)}\cdot{f(x)+{x\over 1+x}\over x}. 
\end{eqnarray}
In some cases it is more convenient to define properties of halos
by using maximum rotational velocity $V_{max} =
\sqrt{(GM/r)}\vert_{max}$, halo parameter which is probably most
easily related to an observable quantity,
instead of the concentration $C$ or the virial mass $\Mvir$. For
profile Eq.(\EqNFW)  the maximum of the rotational velocity occurs at
$r_{max} \approx 2r_s$.  This gives
\begin{eqnarray}
 V^2_{max}   &=&{G\Mvir\over r_s}\cdot {f(2)\over 2f(C)}, \quad
f(2)\approx 0.432\nonumber \\ 
 M(r)    &=&{r_sV^2_{max}\over G}\cdot {2f(x)\over f(2)}, \quad
\omega^2(r) = {V^2_{max}\over r_s^2}
                             {2f(x)\over x^3f(2)}, \\
V^2_{esc}     &=&-2\phi(r) = 4V^2_{max}
                             {\ln(1+x)\over xf(2)}, \nonumber
\end{eqnarray}
where $V_{esc}$ is the escape velocity at the distance $r$
from the cluster center.

The tidal radius, $r_t$, of a small halo with mass $m$ and maximum
rotational velocity $v_{max}$ moving at a radius $R$ from the center of
a large halo with mass $M(R)$ and $V_{max}$, is the minimum of two
radii: (1) a radius at which the gravity force of the small halo
$F_{grav}$ is equal to the tidal force of the large halo $F_{tide}$,
and (2) a radius defined by the resonances between the force the small
halo exerts on the particle and the tidal force by the large halo. If
$r$ is the distance of a particle from the center of the small halo,
then the condition $F_{grav}(r) =F_{tide}(r;R)$ gives an equation for
the tidal radius $r_t$:

$$
\left(R\over r_t\right)^3{m(r_t)\over M(R)} =
         2-{R\over M}{\partial M \over \partial R}, 
$$
\begin{equation}
{f(x_r)\over f(x_R)} =
            \left({x_r\over x_R}   \right)^3
            \left({r_sV_{max}\over R_sv_{max}}\right)^2
       \left( 2- {x_R^2\over (1+x_R)^2f(x_R)}\right),   
\end{equation}
where $x_r \equiv r_t/r_s$ and $x_R \equiv R/R_s$. The last equation
can be solved numerically.

It was argued (e.g., Weinberg 1994ab, 1997) that effective tidal
stripping can occur at smaller radius defined by resonances between
the force the small halo exerts on the particle and the tidal force by
the large halo. We assume that stripping mainly happens at primary
resonance $\omega(r)\vert_{small} =\omega(R)\vert_{large}$. This leads to the
following equation for the tidal radius:
\begin{equation}
{f(x_r)\over f(x_R)} =
            \left({x_r\over x_R}   \right)^3
            \left({r_sV_{max}\over R_sv_{max}}\right)^2,
\end{equation}
 We take the smaller of the two estimates of $r_t$.
For $x_R>2.2$ the tidal radius is defined by the equal force
condition. At smaller distances the orbital-internal resonance defines
the tidal radius.

\begin{figure*}[ht]
\pspicture(0,0)(18.5,18.5)

\rput[tl]{0}(1.,19.5){\epsfxsize=17cm
\epsffile{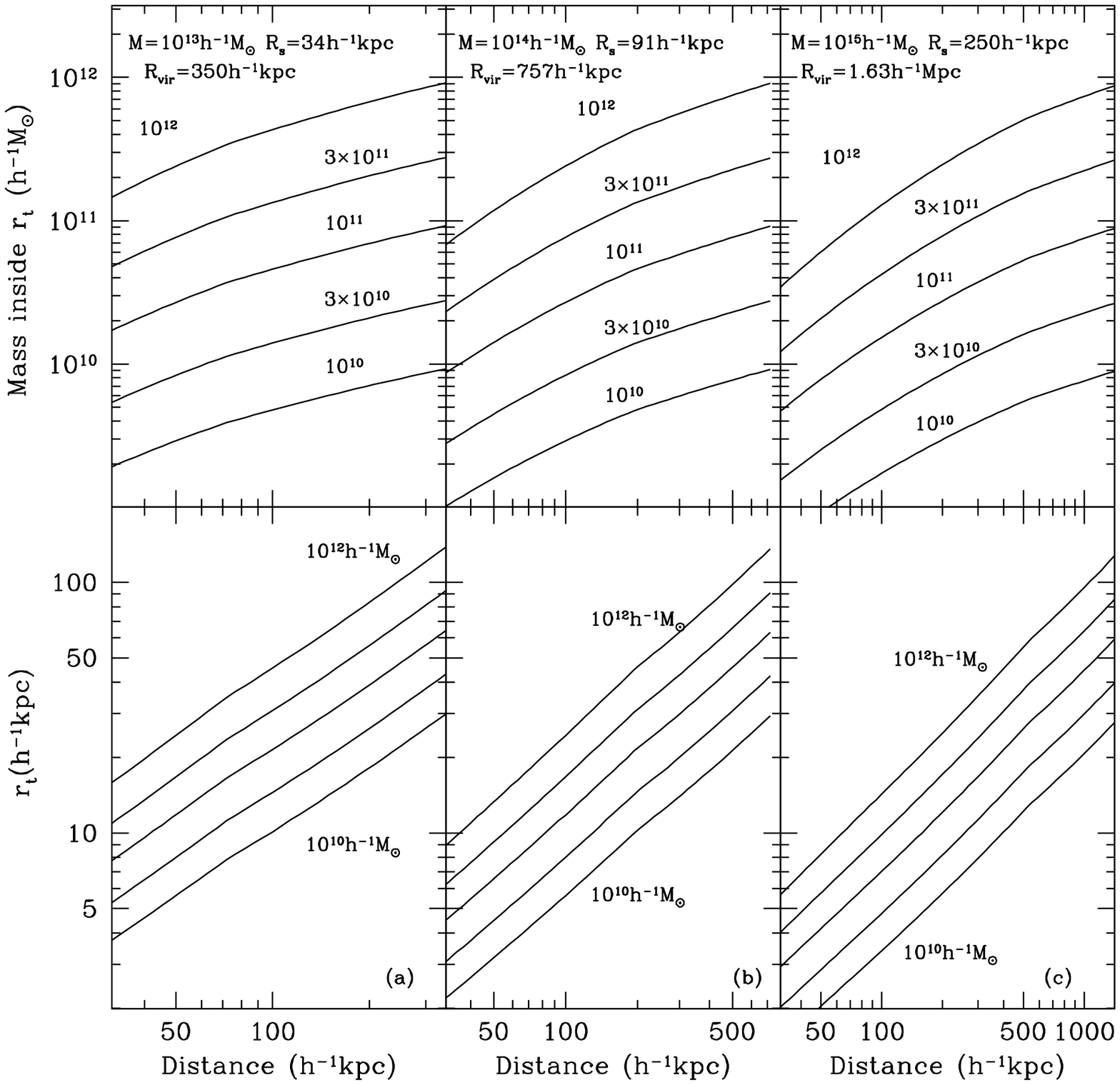}}

\rput[tl]{0}(0.,1.5){
\begin{minipage}{18.5cm}
  \small\parindent=3.5mm {\sc Fig.}~1.--- Tidal radii (bottom row) and
  masses within the tidal radius (upper row) for halos at a given
  distance from the center of a galaxy cluster of mass $10^{13}\Mhsun$
  (a), $10^{14}\Mhsun$ (b), and $10^{15}\Mhsun$ (c). The density
  profiles for both the clusters and the halos are given by Eq.(\EqNFW)
  with an appropriate concentration $C(M)$.  In the figures the mass of
  a dark halo, indicated next to each curve, is the mass $\Mvir$ of the
  halo {\em before} tidal stripping, when the halo was outside the
  cluster virial radius.
\end{minipage}
}
\endpspicture
\end{figure*}

Figure {\Tideone} shows tidal radius and mass within the tidal radius
for halos at a given distance from the center of a group of galaxies
with different masses. In this figure the mass of a dark halo,
indicated next to each curve, is the virial mass of the unstripped
halo, i.e. mass {\em before} the halo entered the cluster. As the
distance from the cluster center decreases, the tidal radius of the
halo and the mass within the tidal radius decrease accordingly.  Even
at large distances from the cluster center ($R>200\kpch$) the halo
radius changes significantly. For example, a halo with
$\Mvir=10^{12}\Mhsun$ and $\Rvir=163\kpch$ at $R=200\kpch$ from the
center of $10^{13}\Mhsun$ group lost only 20\% of its original mass,
but its radius decreased by a factor of two.  The mass inside the tidal
radius $m(r_t)$ changes with $R$ much slower $m\propto R^{0.3-0.5}$
than it would for the isothermal distribution for which we expect
$m\propto R$. This is because the halos with profile Eq.(\EqNFW) are
more centrally concentrated $\rho_{halo}\propto R^{-3}$ than isothermal
halos ($\rho_{iso}\propto R^{-2}$).  Note that the central cusp in
Eq.(\EqNFW) is not important for survival of halos at large distances:
$r_s$ is smaller than the tidal radius $r_t$.  At smaller distances
($R\lesssim 2.2R_s$) mass within the tidal radius decreases faster
($m\propto R$) and the orbital-internal resonance defines the tidal
stripping.  It is likely that we overestimate the effect of tidal
destruction at these distances as compared to galaxies in real clusters
because the tidal radius is small $\sim 10\kpch$ and the baryonic
component cannot be neglected. At the same time, whether halos survive
or not, they already have lost a very large fraction of their mass
($\sim 90\%$, exact number depends on parameters of the halo) when they
get to $R\lesssim R_s$.

\subsection{Effects of energy dissipation by baryons}

Baryonic matter can loose energy by emitting radiation. This
dissipation allows baryons to sink onto the centers of halos and
produce a dense central region inside parent dark matter halo. Because
a denser halo is more resistant to the tidal disruption, baryon
dissipation clearly helps galaxies to survive in clusters. We consider
two effects due to the energy dissipation. (i) The shape of the
density profile in the central part of the halo changes without
changing much overall structure of the halo. (ii) Global parameters of
the halo (such as its characteristic radius $R_s$ and maximum
rotational velocity $V_{max}$) change in reaction to the motion of
baryons into the center. Another possible effect is orbit 
circularization due to formation of rotationally supported baryonic
disk. The effect of this last process, however, is
difficult to estimate, but it likely affects only baryons. Below we
discuss the first two effects.

(i) {\it The shape of the density profile in the halo center.}
The main question here is how much our estimates of the tidal radius
would change if we assumed a steeper central cusp. 
\newpage
{\pspicture(0.5,-1.5)(13.0,13.0)
\rput[tl]{0}(-1.,13.6){\epsfxsize=10.5cm
\epsffile{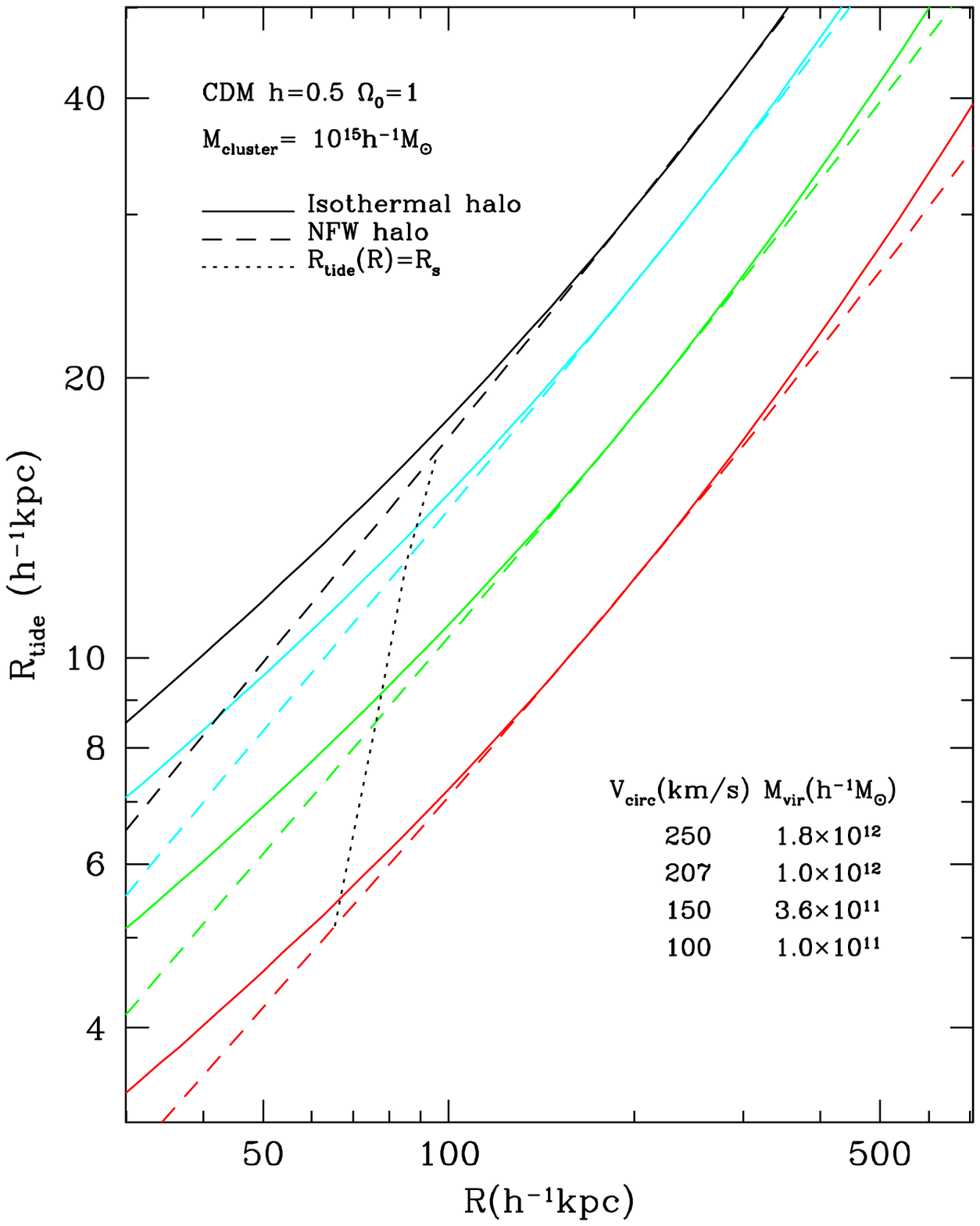}}
\rput[tl]{0}(0.1,3.5){
\begin{minipage}{8.7cm}
  \small\parindent=3.5mm {\sc Fig.}~2.--- Tidal radii of
  halos of different masses and circular velocities inside a cluster
  of virial mass $10^{15}\Mhsun$. Isothermal halos (solid curves) have
  larger tidal radii as compared with the NFW halos (dashed curves),
  which means that isothermal halos are more stable.  Both isothermal
  and NFW halos have the same maximum circular velocity.  The dotted
  curve shows distance at which the tidal radius is equal to the
  characteristic radius $R_s$. On the left from the curve the halos are
  tidally destroyed no matter what resolution is used. While the energy
  dissipation and subsequent baryonic infall increase the tidal radius,
  realistic dissipation is hardly a dramatic effect for halos.
\end{minipage}
}
\endpspicture}
To get an estimate
of the effect, we assume that sinking of the baryons to the halo center
produces flat rotation curve: $V_c=$constant, $\rho\propto r^{-2}$. 
For simplicity, the rotational velocity is assumed to be equal to the
maximum of the circular velocity of the DM halo before the dissipation.
Figure {\TideIso} shows tidal radii of halos of different masses and
circular velocities inside a cluster of mass $10^{15}\Mhsun$.
Isothermal halos have larger tidal radii, which means that they are
more stable. Nevertheless, the difference with the NFW halos is very
small at distances from cluster center larger than $\sim 50\kpch$. Even
at $(30-50)\kpch$ the the difference is only 10\%-20\%.  The dotted
curve in this plot shows the distance at which the tidal radius is
equal to the characteristic radius $R_s$ of the satellite halos. At this
distance the tidal radius of the isothermal halos is $\lesssim 5\%$
larger than tidal radius of the NFW halos. Our $N$-body simulations
presented in the next section indicate that halos which come that close
to the center of a cluster will be tidally destroyed regardless of what
resolution is used.  They may leave a very small leftover, which, even
if present, is so small that it cannot represent a galaxy. Thus, the
change in the shape of the central density profile cannot significantly
affect survival of halos inside clusters.

(ii) {\it Increase of the central density due to dissipation}.
The next question is what effect for halo survival can have possible
increase of the central density (and hence maximum rotational velocity)
due to the baryonic dissipation. Again, we assume that the halo after
the dissipation has a flat rotation curve and, thus, its tidal radius
can be roughly estimated using isothermal density profile. In Figure
{\TideIso} the halo ``moves'' from the NFW curve (dashed line) to a
higher isothermal curve (solid line). For example, a DM halo with the
virial mass of $10^{11}\Mpch$ with tidal radius of $7\kpch$ at
$100\kpch$ from the cluster center can increase its tidal radius to
$10\kpch$, if its circular velocity increases from $100\kms$ to
$150\kms$. This may save the halo from the tidal destruction because
its tidal radius is now about twice larger than its $R_s$.
Unfortunately, there are limits on the increase of the rotational
velocity. The gas possesses non-zero angular momentum, characterized by
the dimensionless spin parameter $\lambda$ (Binney \& Tremain 1987) and
sooner or later becomes rotationally supported.  For a typical value of
$\lambda=0.05$, the adiabatical infall model (e.g., Mo, Mao, \& White
1998) predicts an increase of the maximum rotational velocity of $V_{c
  disk}/V_{c halo}=1.3$. (We use eqs. 33-34 in Mo, Mao, \& White with
the fraction of mass in disk $m_d=0.05$; a correction is made to
convert from $V_{200}$ to $V_{max}$).  Using a different approach to
treat the baryonic infall, Avila-Reese, Firmani, \& Hernandez (1998)
arrive at the same correction: rotational velocity increases by a
factor 1.3-1.4 for typical halo parameters.  Figure {\TideIso} shows
that such increase would correspond to $\sim 20-40\%$ increase in the
tidal radius. While the energy dissipation and subsequent baryonic
infall increase the tidal radius, it is hardly a dramatic effect for
realistic halos.

\subsection{Tidal disruption of halos: simple numerical simulations}

Up to this point we have treated the tidal stripping in a rather
simplified way. Such simplified treatment is useful, because it gives a
rough approximation for a complicated process. In reality (even within
the framework of dark matter dynamics), the situation is more
complicated. This is especially true for halos that loose a significant
fraction of mass, in which case stability of the halos against the
two-body relaxation is an issue.  When a halo looses mass, the mass is
lost from peripheral parts. If the trajectories of particles are not
circular (which is typically the case), the central region of the halo
will start to ``feel'' the loss on a dynamical time-scale. Some of
particles leave the center and are not replaced by tidally-stripped
particles; this decreases the central density and leads to the
expansion of the halo and to further loss of mass. The cycle may
repeat, destroying eventually the entire halo. Whether this process
leads to halo destruction or not depends on the orbit of the halo and
its tidal radius.

In order to study the tidal stripping in a more realistic way, we run a
set of small $N-$body simulations in which a halo of a few thousand
self-gravitating particles moves in a rigid potential of a cluster. The
cluster of the virial mass $2\times 10^{14}\Mhsun$, a typical cluster
mass for large-scale simulations presented later in the paper, is
assumed to have the NFW density profile eq.(\EqNFW). To be consistent
with our large simulations, we also use a $\Lambda$CDM cosmological
model with the following parameters: $\Omega_0=1-\Omega_{\Lambda}=0.3$,
$h=0.7$, $\sigma_8=1.0$. The halo mass is chosen to be $10^{12}\Mhsun$.
This halo has the characteristic radius of $R_s=19.5\kpch$ and the maximum
circular velocity of $190\kms$.  
{\pspicture(0.5,-1.5)(13.0,15.0)
\rput[tl]{0}(-1.25,15.3){\epsfxsize=12.5cm
\epsffile{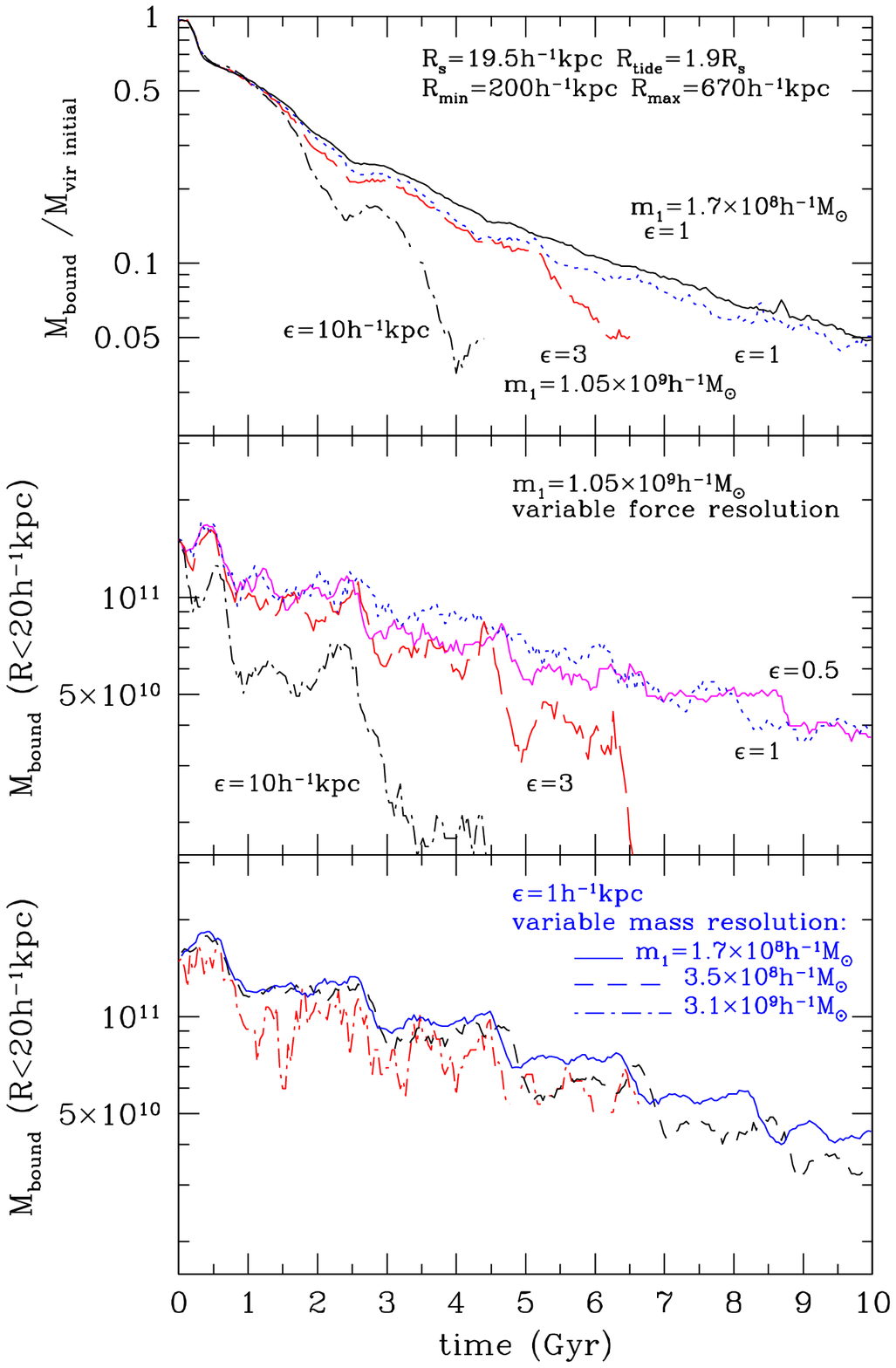}}
\rput[tl]{0}(0.5,2.9){
\begin{minipage}{8.7cm}
  \small\parindent=3.5mm {\sc Fig.}~3.--- The evolution of the total
  bound mass (top panel) and the bound mass inside central $20\kpch$
  radius (middle and bottom panels) for a halo with initial mass
  $10^{12}\Mhsun$ moving inside a cluster of virial mass $2\times
  10^{14}\Mhsun$. Different curves correspond to different force and/or
  mass resolutions used in the experiment (each curve is marked with
  corresponding resolution).  Both the cluster and (initially) the halo
  have the NFW density profiles. The simulations were done with
  different mass $m_1$ and force $\epsilon$ resolution. The halo looses
  95\% of its mass, but given sufficient resolution it survives for
  more then 10Gyr.
\end{minipage}
}
\endpspicture}
Within the radius of
$R_{200}=163\kpch$ the halo has overdensity 200 relative to the
critical density of the Universe.
At the initial moment the particles of the halo were distributed
inside $R_{200}$ in 100 equally spaced spherical shells in
such a way that the density profile obeys eq.(\EqNFW). Eq.(\EqVelDisp)
was used to find one dimensional velocity dispersion for each
shell. The velocity dispersion was used to assign velocities to
individual particles by  throwing three random gaussian numbers for
each velocity vector. The velocity distribution was isotropic. This procedure
generates a system with the NFW profile which is almost in
equilibrium. The shot noise (the finite number of particles) results in
residual deviations from the equilibrium. The finite extent of the
system also produces transient effects at the outer boundary.  By
running an isolated system with 3000 particles for a long time (five
billion years) we tested that the system is really close to a stationary
NFW halo. We found only one deviation. The density distribution at the
outer boundary (within 30\% of $R_{200}$) was slightly smeared
out. 
{\pspicture(0.5,-1.5)(13.0,11.5)
\rput[tl]{0}(-.5,12.5){\epsfxsize=10.5cm
\epsffile{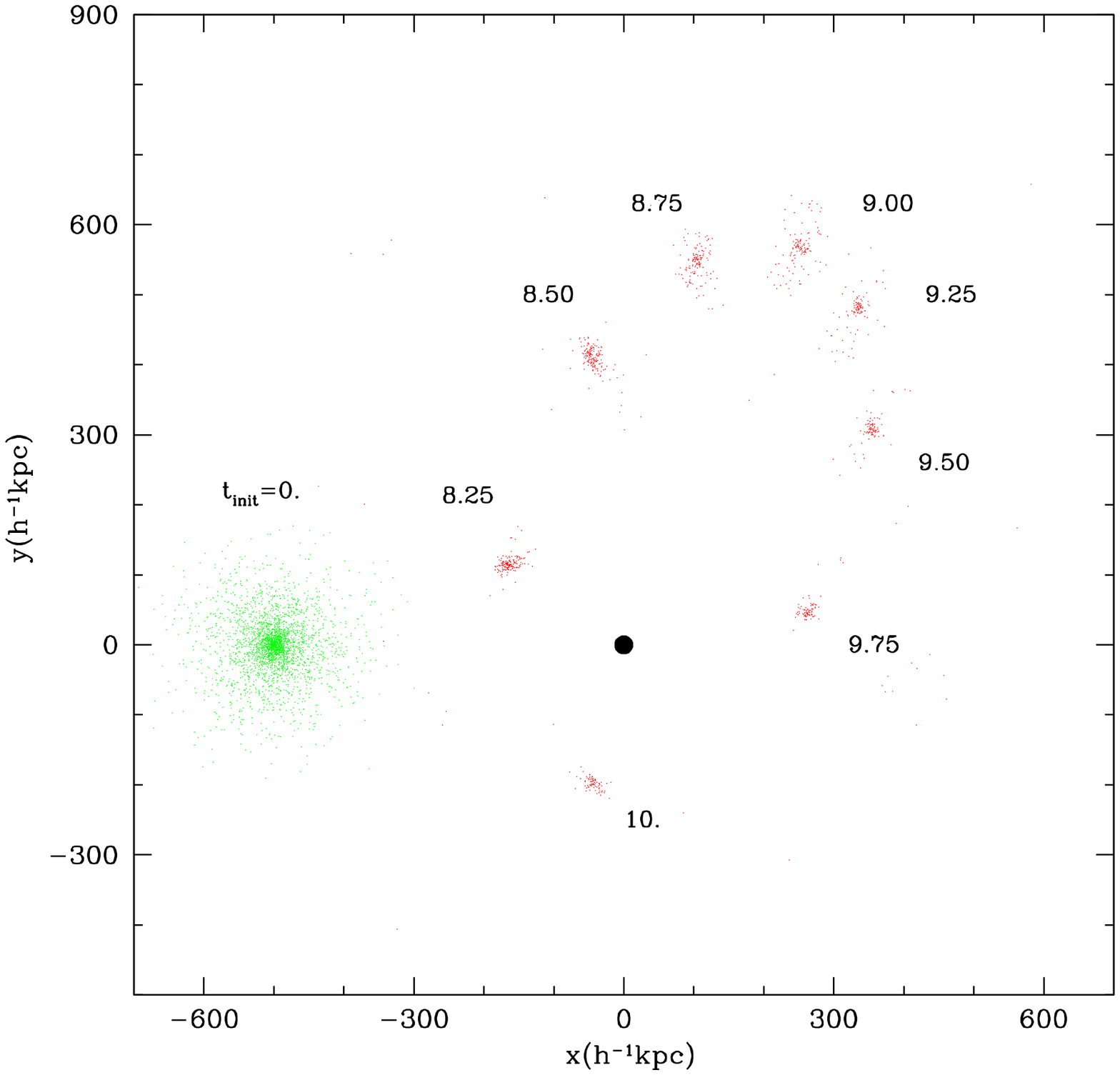}}
\rput[tl]{0}(.5,2.){
\begin{minipage}{8.7cm}
  \small\parindent=3.5mm {\sc Fig.}~4.--- Positions of bound particles
  for halo presented in Figure {\TideResolution}. The time (marked next
  to the halo) is given in billions of years.  The solid circle
  indicates position of the cluster center. The halo initially had 952
  particles and has lost most of its mass after 10~Gyrs. Nevertheless,
  even after 10Gyr, it can still be identified as a compact centrally
  concentrated bound lump of particles.
\end{minipage}
}
\endpspicture}
With the exception of the shot noise, no deviations were found
inside 70\% of $R_{200}$.

We use the Aarseth's N-body presented in Binney \& Tremaine (1987). The code
was modified to include the acceleration from the cluster. The code
uses Plummer softening of the gravitational potential. The softening is
thus parameterized by a softening length $\epsilon$. It should be noted that
the  Plummer model gives a softer force as compared with the ART code,
which we used for large-scale simulations. We estimate that the
difference in resolution is about factor of two: 10\% error in
acceleration is reached at $\sim 2$ cells in ART code as compared with
$\sim 3.7\epsilon$ in a code with the Plummer softening (Kravtsov et al. 
1997).

As a typical example, we have chosen an elliptical orbit with the
minimum distance to the cluster center of $200\kpch$ and ratio of
maximum to minimum distances of 3.35. Initially the halo was placed at
distance $500\kpch$ and was given velocity $710\kms$ (circular velocity
at that distance) in the direction of 45 degrees to the radius. The
period of the resulting orbit is $2\times 10^9$yrs.  The tidal radius
at pericenter was $37\kpch\approx 1.9R_s$ and mass within the tidal
radius was $1/3$ of the halo's initial virial mass virial mass.  We
study the effects of the force and mass resolution on the evolution of
the halo varying the resolutions by a factor of 20.

Figure {\TideResolution} shows the evolution of the total bound mass
(top panel) and the bound mass inside central $20\kpch$ (middle and
bottom panels). We show the results only if more than 15 bound
particles are 
{\pspicture(0.5,-1.5)(13.0,13.5)
\rput[tl]{0}(-1.,13.5){\epsfxsize=12.5cm
\epsffile{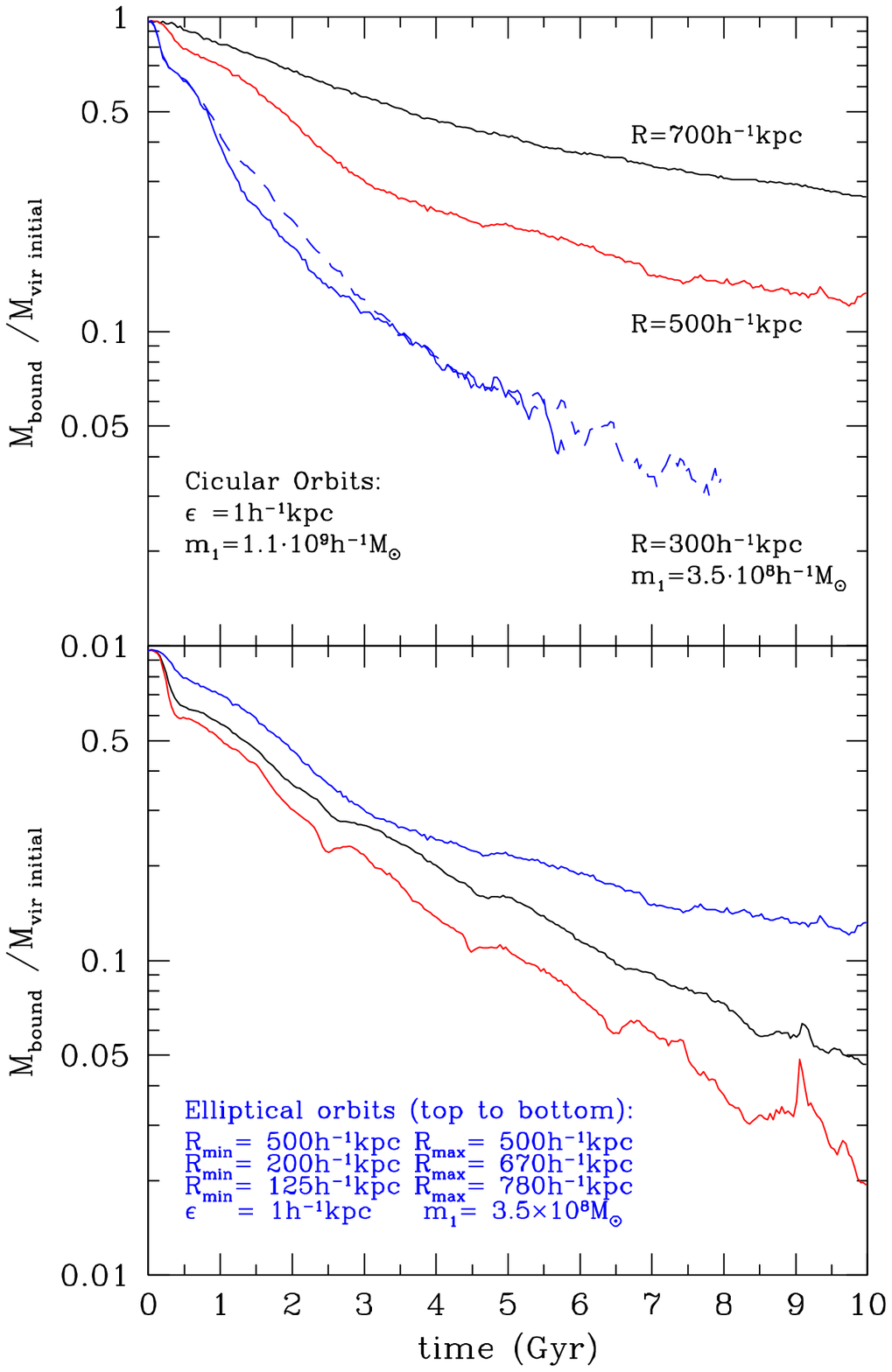}}
\rput[tl]{0}(0.5,1.5){
\begin{minipage}{8.7cm}
  \small\parindent=3.5mm {\sc Fig.}~5.--- Mass evolution for halos on
  different orbits. While halos on the highly eccentric orbits, with
  pericenters as small as $125\kpch$, have survived for at least 10Gyr,
  a halo on a circular orbit with radius $300\kpch$ was tidally
  destroyed after about 7 Gyrs regardless of the used resolution.  Note
  that improving mass resolution by a factor of three allows us to
  track the halo for a longer time. But results indicate convergence at
  mass fractions above 5\%.
\end{minipage}
}
\endpspicture}
found in the central region. The halo is considered
``lost'' if the number of bound particles is less than 15.  For halos
in the middle panel we fixed the mass resolution and studied the effect
of the force resolution. If the force resolution is not sufficient, the
halo is lost. For example, with the force resolution of
$\epsilon=10\kpch\approx R_s/2$ the halo is lost after 4Gyr. Increasing
the force resolution helps the halo to survive, but once a threshold of
$\epsilon=(1-2)\kpch$ is reached, additional increase in resolution
does not change the mass of bound particles, which indicates
convergence of the results. Extra resolution may even reduce the bound
mass because of excessive two-body scattering, which was observed in
some of our simulations.

The bottom panel of Figure {\TideResolution} illustrates the effects of
the mass resolution. In this case the force resolution was fixed to 
$\epsilon =1\kpch$, sufficient for halo survival with mass
resolution $m_1=1.05\times 10^9\Mhsun$ of the previous plot. The halo
with three times worse mass resolution was lost after 6.5Gyr. As in the
case of varying force resolution, the increase of mass resolution above
some limit does not result in increase of the bound mass, which again
points to convergence. This can be seen more clearly in the top panel: when
the mass resolution is improved by a factor of 6 (from the dotted to
the solid curve), the mass of bound 
{\pspicture(0.5,-1.5)(13.0,16.5)
\rput[tl]{0}(-2.,16.5){\epsfxsize=13.5cm
\epsffile{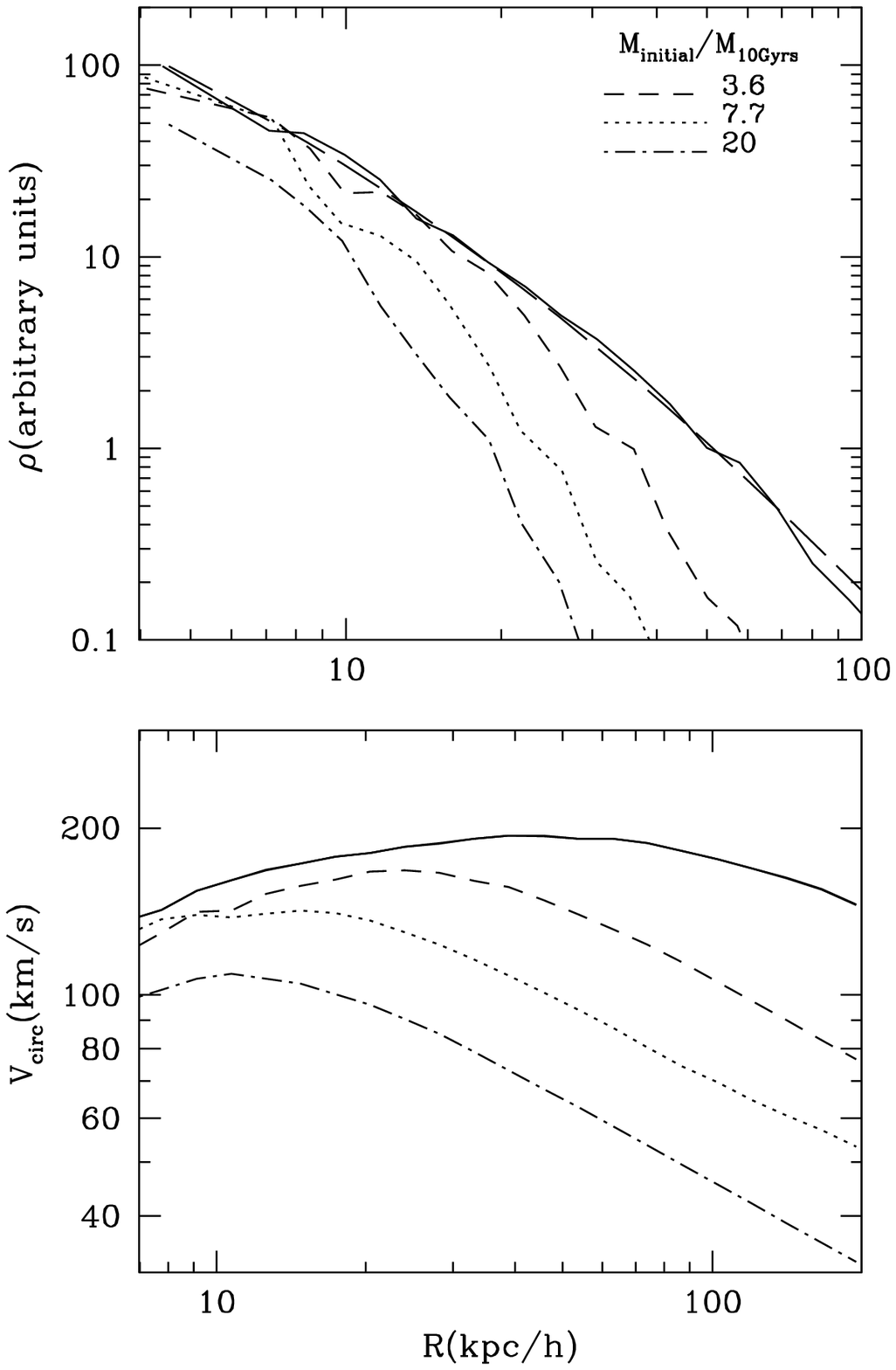}}
\rput[tl]{0}(0.5,3.5){
\begin{minipage}{8.7cm}
  \small\parindent=3.5mm {\sc Fig.}~6.--- The evolution of density
  (top) and velocity (bottom) profiles for a halo of mass
  $10^{12}h^{-1}{\ }M_{\odot}$ orbiting inside a $2\times
  10^{14}h^{-1}{\ }M_{\odot}$ cluster.  In both panels, the {\em solid}
  curves show initial profile; the {\em long dashed} curve in the top
  panel shows the NFW profile for the initial distribution; {\em
    short-dashed}, {\em dotted}, and {\em dot-dashed} curves show
  profiles for halos on different orbits after 10~Gyrs of orbiting in
  the cluster. The orbits of the halos are the same as the two top
  curves in panels of Figure {\TideOrbits} (one orbit was shown twice).
  The curves correspond to different initial to final mass ratios
  (shown in the legend). Note that the central $\sim 10{\ }{\rm kpc}$
  region is less affected by the tidal stripping, but when the halo
  looses more than $90\%$ of its mass even the center becomes affected
  noticeably.
\end{minipage}
}
\endpspicture}
particles changes only by 10\%-15\%.

It should be stressed that the mass loss of the halo was really
dramatic: after 10Gyr the halo was 20 times less massive than at the
beginning. The mass of the whole halo is a function of time, which
steeply declines in the beginning and tends to level out at later
moments. The mass loss from the central region behaves as a step
function: after every passage near the cluster center the mass
drops by 20\% -- 30\%.  Nevertheless, the halo has survived (i.e., was
detectable) for at least 10Gyr. Figure {\TideOrbit} illustrates the
dramatic mass loss of the halo. In this plot we show only bound
particles for a halo which initially had 952 particles. While the halo
has lost most of its mass, even after 10~Gyr it is still a compact
centrally concentrated dense lump of particles.

Survival or destruction of halos depends on parameters of their orbits
in the cluster. To study the orbit dependence, we have run similar
simulations varying orbit eccentricities and radii. We have found that
in general halos on circular orbits suffer larger mass loss than halos
on the eccentric orbits. Although the pericenter of the latter may be
smaller, the halos spend only a small fraction of the orbital period in
the central dense region of cluster.  Figure {\TideOrbits} shows mass
evolution for halos on different orbits. While halos on the highly
eccentric orbits, with pericenters as small as $125\kpch$, have
survived for at least 10Gyr, a halo on a circular orbit with a radius
of $300\kpch$ was tidally destroyed after $\approx 7$ Gyrs, regardless
of the used resolution. 

Finally, Figure {\DVevol} shows evolution of the density and circular velocity
($V_c(r)=\sqrt{GM(<r)/r}$) profiles of the halo for the same orbits as shown in
Figure {\TideOrbits} during a 10 Gyr period of orbital evolution.
Different orbits lead to different mass loss rates and the figure shows
the cases in which the ratio of initial to final bound halo mass is
$3.5$, $7.7$, and $20$. The figure shows that significant mass loss
results in dramatic changes of the density and velocity profiles at
large radii. The changes, however, are not as dramatic for the central halo
regions. Nevertheless, in the case of extreme mass loss ($\approx
95\%$) shown by the dot-dashed curves, both density and maximum
circular velocity decrease by a factor of two from their initial
values.

\subsection{Erasure of substructure via dynamical friction}

The dynamical friction is another effect that contributes to the
erasure of the substructure. This is not a numerical effect, but by
driving galaxies to the cluster center where they will be tidally
destroyed, it can enhance numerical effects.  The dynamical friction
time for a small halo with mass $m$ moving on a circular orbit of
radius $R$ around the large halo can be estimated using the
Chandrasekhar's formula (Binney\& Tremaine 1987) with assumptions of
equilibrium and a Maxwellian isotropic distribution of velocities of
the DM particles.  The rate of the orbital radius decay due to the
dynamical friction is given by
\begin{eqnarray}
{dR\over dt} &=&-{2R\over t_{fric}}\left(\frac{\partial \ln
    M(R)}{\partial \ln R}+1\right),\nonumber\\
         t_{fric} &=&{V^3_{circ}\over 4\pi G^2(\ln \Lambda) m(r_t) \rho(R)
                [{\rm erf}(X)- 2X e^{-X^2}/\sqrt{\pi}]}, \nonumber\\
V^2_{circ} &=& GM(R)/R, \quad X={ V_{circ}\over \sqrt{2} \sigma_r }\\
\ln \Lambda &=& \ln\left({R_{vir}\over R}
                               {M(R)\over m(r_t)}\right) ,\nonumber\\
M(R) &=& -{R\sigma^2_r\over G}\left[{d\ln\rho\over d\ln R} +
           {d\ln\sigma^2_r\over d\ln R}
                           \right].\nonumber
\end{eqnarray}
For the density profile Eq.(\EqNFW) we have 
$d\ln\rho/ d\ln R =-(1+3x)/(1+x)$, where $x=R/R_s$. The last of
the equations (\EqDynFr) can be rewritten as an equation for the
velocity dispersion:
\begin{equation}
{d\sigma^2_r\over dx} -{1+3x\over x(1+x)}\sigma^2_r =
              - {G\Mvir\over R_sf(C)}\cdot {f(x)\over x^2}
\end{equation}
The solution of the equation is
\begin{eqnarray}
\sigma^2_r &=& {x(1+x)^2\over 4}\left[
                            \sigma^2_0 - 4{G\Mvir\over R_sf(C)}
                          \int^{x}_1{f(x)\over x^3(1+x)^2}dx
                                                           \right] \nonumber\\
                     &=& V_{max}^2{2x(1+x)^2\over f(2)}\int^{\infty}_x{f(x)\over x^3(1+x)^2}dx
                           \\
\sigma^2_0 &=& 4{G\Mvir\over R_sf(C)}
                            \int^{\infty}_1{f(x)\over x^3(1+x)^2}dx
                            \approx 0.432V^2_{max}.\nonumber
\end{eqnarray}
Here $\sigma_0$ is the 1D velocity dispersion at $R=R_s$. The
velocity dispersion $\sigma_r$ has a maximum $\sigma_r\approx \sigma_0$
at $R\approx 0.8R_s$. It declines on smaller and larger radii, but the
maximum is  flat: $\sigma_r\approx 0.78\sigma_0$ at $R=
0.1R_s$ and $\sigma_r\approx0.69\sigma_0$ at $R=\Rvir$.  Equations
(\EqDynFr) and (\EqVelDisp) define the dynamical friction time, which
is presented in Figure {\Dynfriction} for different masses of
clusters and halos. For a given halo, the dynamical friction time
decreases as the halo moves into the cluster because the density of
cluster increases. Note, however, that this decrease is countered by 
the halo mass decrease due to the tidal stripping. The dynamical
friction time is thus a varying quantity which depends on the cluster
mass, distance from the cluster center, and halo's gravitationally
bound mass. 

We should note that equation {\EqDynFr} should be considered as a rough
estimate of the effect for at least two reasons. First, we assume for
simplicity that orbits are circular, which, of course, is a very
simplified view of the real halo orbits in clusters.  This
simplification, however, makes it possible to estimate all the
interesting quantities analytically and thus greatly facilitates
computations of dynamical evolution. There is very little data as to what
kinds of orbits are to be expected (see, however, recent work of van
den Bosch et al. 1998). Recent numerical studies by Tormen
(1997), Tormen et al (1998), and Ghigna et al. (1998) give somewhat
different accounts on the distribution of the orbital parameters. Tormen
(1997) concludes that orbits are neither radial nor circular and that
the average eccentricity of the orbits is $\epsilon \approx 0.5$, while
orbits in cluster analyzed by Ghigna et al. (1998) are reported to be
mostly radial. It is not clear at present what causes the difference.

Importance of the dynamical friction for a particular halo will depend
on the halo's orbit.  Halos on circular or low-eccentricity orbits with
large radius will experience little gravitational drag and would
never come close to the center and get tidally disrupted.  On an
eccentric orbit, some of these objects would have a chance to pass
though the dense cluster core where the tidal effects discussed in the
previous section (and gravitational drag too) would be considerably
stronger. Note, however, that the time spent by a halo on highly
eccentric orbit near the orbit pericenter is rather small. The strong
tidal force in the cluster center will shock the halo during pericenter
passages, but for most of its orbit the halo will exist in less dense
(and thus more favorable for its survival) environments. The worst
evolution scenario can be envisioned for a halo on a low-eccentricity
orbit with a small radius. Such halo will suffer both strong and
continuous tidal losses and fast decay of the orbital radius due to the
dynamical friction.

The validity of the Chandrasekhar's formula in
spherically symmetric, finite-size systems was also a subject of extensive
debate over the years.  One
obvious drawback of this equation is that it does not predict a drag
experienced by a satellite on an orbit outside of the spherical system
(e.g., Lin \& Tremaine 1983).  Most studies, however, indicate that the
equation works remarkably well. Lin \& Tremaine (1983), for example,
have addressed the subject numerically and have found that eq.
{\Eqtfric} works well for satellites of mass $\lesssim 0.05$ of the
primary. The same conclusion was reached in other numerical studies
(Bontekoe \& van Albada 1987; Zaritsky \& White 1988; or more recently,
Cora et al. 1997).  Analytical study by Weinberg (1986), in which
Chandrasekhar's formula was rederived for spherically symmetric
potential, also shows that this formalism is applicable in most cases.
Recent analytical studies by Bekenstein \& Maoz (1992), Maoz (1993),
and Dom\`{\i}nguez-Tenreiro \& G\'omez-Flechoso (1998) used
fluctuation-dissipation approach to derive the energy losses due to the
gravitational drag. Despite the more sophisticated treatment and
inclusion of the various important details (satellite's size, not
negligible mass of the background particles, inner velocity structure
of satellite), it appears that Chandrasekhar's formula gives a good
approximation for the energy losses.  

Dom\'{\i}nguez-Tenreiro \& G\'omez-Flechoso (1998), for example, have
taken into account both the finite size of the satellite and the
internal velocity structure of the satellite.  Their analysis shows
that if internal velocity dispersion of the satellite is $\lesssim 0.5$
of the velocity dispersion of the primary system (always the case in
our study of massive clusters) the energy losses predicted by
Chandrasekhar's formula are accurate to within $\sim 30\%$. 

There is, of course, a considerably larger uncertainty caused by the
choice of the Coulomb logarithm, because there is no general
prescription of how to estimate the minimum and maximum impact
parameters.  The value of the Coulomb logarithm in clusters of galaxies
is expected to be $\ln \Lambda \approx 5-10$. The value of $\ln
\Lambda=8$ was used in a recent study by Tormen et al.  (1998), who
compared dynamics of dark matter halos in clusters in the cosmological
context. They showed that even when $\ln \Lambda$ is kept constant the
Chandrasekhar's formula works remarkably well, predicting quite
accurately decay of the satellite's orbital radius for the satellites
as massive as $\sim 0.2-0.5$ of the cluster mass. This latter test is
most relevant to our study and shows that use of the eq.  {\Eqtfric} is
justified.

The dynamical friction time as a function of the distance to the
cluster center for a variety of cluster and satellite masses is shown
in figure {\Dynfriction}.  In general, the results are hardly
surprising. Dynamical friction in rich clusters $\Mvir\approx
10^{15}\Mhsun$ is negligible except for the most massive galaxies near
the cluster center. For poor clusters and groups with $\Mvir \lesssim
10^{14}\Mhsun$, the friction time is short as compared with the Hubble
time. Adding the baryons would only shorten the friction time, as more
mass would be retained within the halo's tidal radius. If a group
exists for a sufficiently long time and does not accrete efficiently
new satellites, the dynamical friction would produce an object that
would look like an overmerger -- a giant central galaxy with no other
galaxies in the group\footnote{Notably, such systems are
  observed in the real Universe (Carignan et al. 1997).}. The epoch of
formation and the growth rate of groups depends on the parameters of a
cosmological model. Thus, if resolution is sufficient to make the
numerical effects negligible, excessive overmerging for groups and poor
clusters should indicate that the cosmological model is wrong.

{\pspicture(0.5,-1.5)(13.0,13.5)
\rput[tl]{0}(-1.25,13.5){\epsfxsize=11.5cm
\epsffile{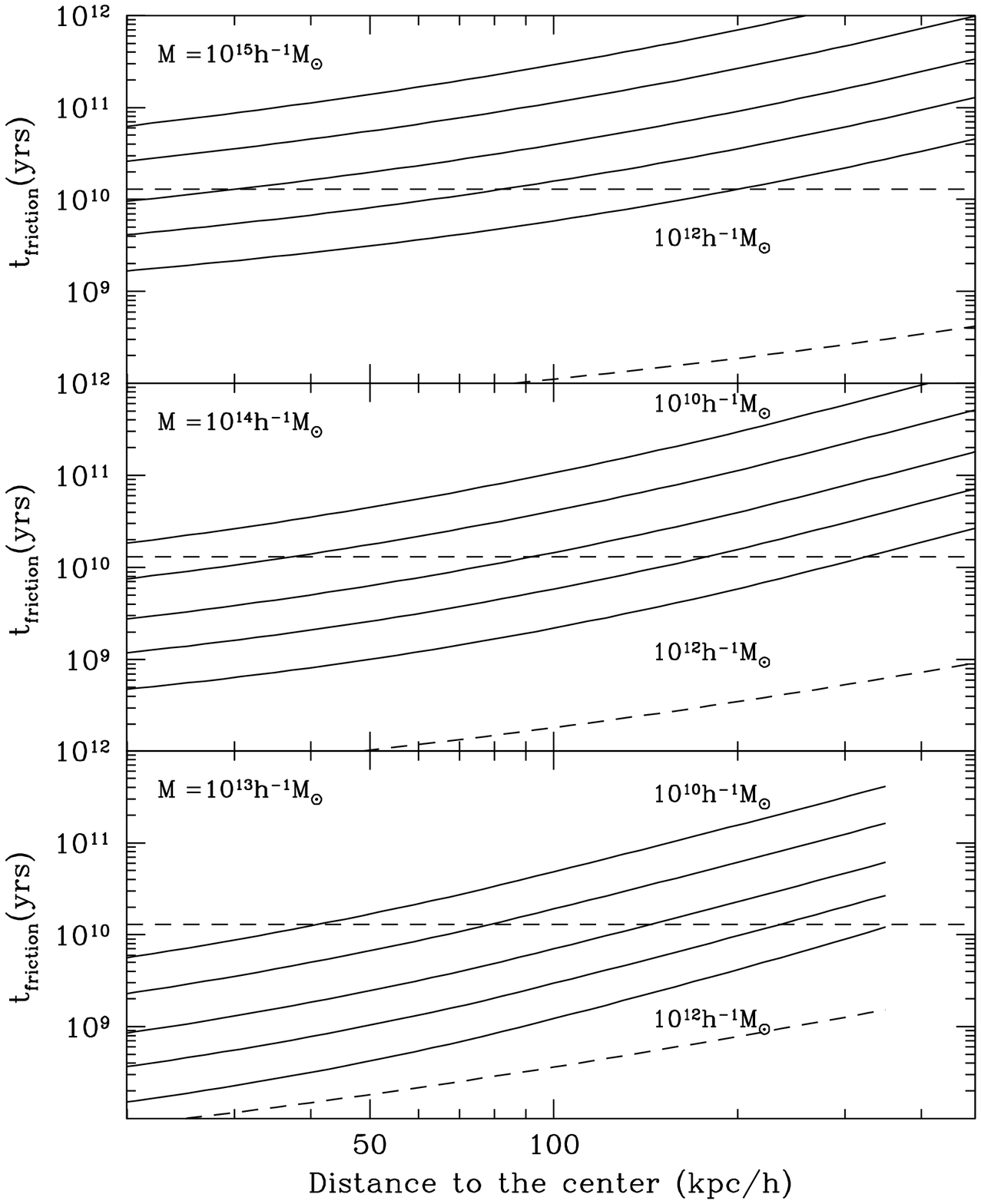}}
\rput[tl]{0}(0.,1.5){
\begin{minipage}{8.7cm}
  \small\parindent=3.5mm {\sc Fig.}~7.--- The dynamical friction time
  for different masses of clusters and halos. For a given halo the
  dynamical friction time decreases as the halo moves into the cluster
  because the density of cluster increases. But then the halo starts to
  lose its mass, and the friction time increases again. The set of
  virial masses is the same as in Figure \Tideone. The horizontal
  dashed line shows the Hubble time for this cosmological model.
\end{minipage}
}
\endpspicture}
A matter of concern is how one could distinguish between numerical and
physical effects as the primary cause of overmerging. There may be two
approaches to this problem.  First, one can estimate the required
resolution in the manner outlined in the next section and run a
simulation.  After that run another simulation with considerably higher
resolution.  Such a test may be prohibitive in terms of the CPU time
required to run the very high-resolution run.  However, it is probably
the most certain way of testing for the numerical effects.
High-resolution simulations of individual clusters are probably best
suited for this kind of test. Indeed, a similar test was attempted by
Ghigna et al. (1998), who analyzed two runs of the same clusters with
different spatial resolutions. The second possible solution is to
identify cluster's progenitor at an earlier epoch (e.g., $z=1$) and
follow evolution of a sample of test halos in the representative mass
range. This analysis may show what mechanism is really at work (e.g.,
dynamical friction may be clearly observed in the orbit decay) and may
allow one to estimate whether the numerical resolution is adequate.

\newpage
\subsection{What numerical resolution is required?}

Numerical experiments presented by Moore et al. (1996) show that halos
become unstable and get quickly disrupted if tidal radius is smaller
than $\sim 2-3$ times the core radius. The force softening, $r_{soft}$,
usually has effect of creating an artificial core in the density
distribution at scales $\lesssim r_{soft}$. Thus, it can be assumed
(optimistically) that halos get disrupted if $r_t\lesssim 2r_{soft}$.
The details of the tidal stripping will, of course, depend
non-trivially on the mass and orbital parameters of halo. However,
tidal radius for a halo can be approximately defined (Johnston 1998;
Ghigna et al. 1998) as the tidal radius defined by equations {\EqTidal}
at the pericenter of the halo orbit. Numerical studies of satellite
dynamics (e.g., Johnston, Hernquist \& Bolte 1996) indicate also that
$\sim 10-30\%$ mass loss can be expected at every pericentric passage,
which agrees well with results of our numerical experiments. This will
have an additional effect if halo orbits long enough to make several
pericentric passages.

Thus, with an adequate mass resolution, dark matter halos should survive
(at least during the first period of their orbit) inside groups of
galaxies as long as their pericentric $r_t$ is twice or larger than
the force resolution.  These halos may lose a large fraction of their mass,
but they should be able to retain their identity even without the
additional baryonic dissipation.  If the required force resolution
cannot be achieved, the baryons need to be included to alleviate the
problem.

Ultimately, survival of a particular halo will depend on both the mass
and the force resolution. In order for two-body evaporation to be
negligible, halos must contain $\gtrsim 30$ particles. This will define
the mass limit for the surviving halos even if the force resolution is
sufficiently high. A very optimistic estimate, thus, would be that a halo
should consist of at least $\sim 20-30$ particles and its tidal radius
should be larger than two resolution elements.

As an example, we consider a rather typical simulation with particle
mass of $10^{10}\Mhsun$ and the resolution of $30\kpch$ (e.g., Gelb \&
Bertschinger 1994; Ma \& Bertschinger 1995; Tormen 1997). With this
resolution one would naively hope to find $\approx 10^{11}\Mhsun$
satellites -- the virial radius of a halo with this mass is more than
twice larger than the force resolution.  The top panel in Figure
\Tideone(b)~ indicates, however, that halos of this initial virial
mass cannot be found inside a cluster of mass $10^{14}\Mhsun$: due
to the insufficient mass resolution they lose so much of their mass
that they are destroyed by the tidal force.  The mass resolution of
these simulations was sufficient to find what is left of a halo with
$3\times 10^{11}\Mhsun$ at distances $\gtrsim 80\kpch$. However, the
bottom panel in Fig. \Tideone(b) shows that tidal radius for
galaxy-size halos is smaller than two resolution elements within
central $\sim 300\kpch$ and thus no such halos should be expected
there. Therefore, lack of the force resolution resulted in erasure
(``overmerging'') of even fairly massive ($\sim 10^{12}\Mhsun$) halos
within the $300\kpch$.

But even a considerably better resolution may not be sufficient if we
deal with a really massive cluster. For example, Carlberg (1994)
simulated a $2.2\times 10^{15}\Mhsun$ cluster with mass
resolution of $2.27\times 10^9\Mhsun$ and the Plummer force softening
of $\epsilon =9.7\kpch$. According to Gelb \& Bertschinger (1994), the
effective resolution for the Plummer force is $2.6\epsilon$. This gives
the resolution $\approx 25\kpch$, which we use as a limit on the tidal
radius of resolved halos. Analysis of tidal radii for a cluster of this
mass shows that due to the insufficient force resolution, no halos
should exist at distances smaller than $\lesssim 290\kpch$. Halos of 
mass $\Mvir < 10^{11}\Mhsun$ should be tidally destroyed at
$R=590\kpch$. This is consistent with what Carlberg (1994) found in his
simulation.

What resolution is required for halos to survive? The above examples
show that the answer depends on the mass of the cluster and on the mass
of the halo one would like to resolve.  The force resolution should be
(significantly) smaller than the minimal tidal radius of a halo
(defined by the orbit's pericenter). For a $10^{14}\Mhsun$ cluster the
tidal radius for a massive halo of mass $M_h\gtrsim 10^{11}\Mhsun$ at a
distance of $R\gtrsim (60-70)\kpch$ is $r_{t}\approx 10\kpch$. The
force resolution should probably be better than $3\kpch$ for such halo
to survive in the cluster. Because the halo at this distance loses
80\%--90\% of its original virial mass, and because realistically one
needs at least 20--30 particles to identify a halo, the particle mass
should be smaller than $\approx 10^9\Mhsun$.  These estimates may be
optimistic for halos that orbit in cluster for several periods due to
continuing mass loss (see \S 2.2) and possible effects of halo-halo
heating (Moore et al. 1996).

To some extent, the answer is clear. In order to allow a galaxy-size
halo to survive, the force resolution must be much smaller than its
tidal radius and the halo must be represented by many particles. For a
typical large galaxy of the mass $10^{11}\Mhsun$ and tidal radius of
$10\hkpc$ in a $\sim 10^{14}-10^{15}\Mhsun$ cluster, the resolution
must be of $(0.5-3)\hkpc$ and particle mass -- $\lesssim 10^9\Mhsun$.
There is a number of further caveats related to analysis of the halo
remnants. The most important of them is the question of whether it is
possible to recover properties of the original halo from the properties
of the remnant.  Unfortunately, the answer to this question is not
clear and various parameters may be affected in different and
complicated ways. Most of the the halos orbit in cluster for periods
less then 10~Gyrs, and although their peripheral parts are being
stripped, the evolution of the central regions is not dramatic (see
Fig. {\DVevol}). 

\section{SIMULATIONS}

We simulate the evolution of $128^3$ cold dark matter particles in two
cosmological models: a flat low-matter density CDM model with
cosmological constant ($\Lambda$CDM; $\Omega_0=1-\Omega_{\Lambda}=0.3$;
$h=0.7$; $\sigma_8=1.0$) and a model with a mixture of cold and hot
dark matter (CHDM; $\Omega_0=1$; $\Omega_{\nu}=0.2$; $h=0.5$;
$\sigma_8=0.7$ ). The CHDM simulation followed trajectories of
additional $2\times 128^3$ hot particles.  The mass fraction of hot
matter, $\Omega_{\nu}$, is equally split between two types of neutrino
(Primack et al.  1995). Both models were normalized to be consistent
with {\sl COBE} DMR observations (Bunn \& White 1997).  Normalization of the
$\Lambda$CDM model is also consistent with observed abundance of galaxy
clusters (Viana \& Liddle 1996), while normalization of the CHDM model
may be slightly higher than suggested by the data (Gross 1997). Both
simulations are initialized with the same set of initial random numbers
in order to reduce effects of the cosmic variance.

To achieve sufficiently high mass resolution, the size of the
simulation box for both the $\Lambda$CDM and the CHDM models is chosen to
be rather small -- $15\Mpch$. To test the effects of the box size, an
additional simulation of $30\Mpch$ box has been run for the
$\Lambda$CDM model.  The mass of a cold particle in the $15\Mpch$ box
simulations (for $30\Mpch$ box particle mass is 8 times larger) is
$m_1=1.33\times 10^8\Mhsun$ for the $\Lambda$CDM model and is $\approx
2.7$ times larger for the CHDM model. If we assume that $\gtrsim 30$
particles are needed to identify a halo, we expect to be able to
identify halos as small as $4\times 10^9\Mhsun$. This corresponds to
$4\times 10^{10}\Mhsun$ halo before tidal stripping, if 90\% of mass
was stripped by the cluster.

The simulations were done using the Adaptive Refinement Tree (ART) code
(Kravtsov et al.  1997). The code used a $256^3$ uniform grid on the
lowest level of resolution and seven levels of refinement for the
$15\Mpch$ box. Each refinement level doubles the resolution. The
seventh refinement level corresponds to the dynamical range of 32,000
and the resolution of {$\approx 0.5\hkpc$}. The $30\Mpch$ run 
had six levels of refinement and its resolution is $\approx 2\hkpc$. The
code refines an individual cell on a given level $L$ if the number of particles
in this cell (as estimated by the cloud-in-cell method)
exceeds some threshold $N_{th}(L)$.  The threshold is $N_{th}=5$ for
high levels and $N_{th}=10$ is set for low levels $L=0,1$ (Kravtsov
et al.  1997). This choice of thresholds ensures that refinements are
introduced only in the regions of high-particle density and prevents
the two-body relaxation effects.  The increase in spatial resolution
corresponding to each successive refinement level is accompanied by
decrease of the integration time-step by a factor of 2.  The
simulations were started at $z_i=30$ when the rms of the density
fluctuations in the simulation box was $\delta\approx 0.27-0.32$.

The dynamic range of the simulation is justified by the following two 
considerations. First, the code integrates the evolution in {\em
  comoving} coordinates. Therefore, to prevent degradation of force 
resolution in {\em physical} coordinates, the dynamic range between the
start and the end ($z=0$) of the simulation should increase by
$(1+z_i)$: i.e., for our simulations $256\times(1+z_i)=7680$. Second, 
the code reaches its peak resolution in the highest density regions 
inside virial radius of the DM halos. To resolve a halo, we need at
least $\sim 10$ resolution elements per halo size, which justifies 
the dynamic range of $\gtrsim 10,000$. 

Particle trajectories were integrated with the step in expansion factor
of $\Delta a_0=0.0015$ on the zero level uniform grid, and with time
step $\Delta a_L=\Delta a_0/2^L$ on a refinement level $L$. This gives
an effective number of steps of 82,000 on the seventh level of
refinement. In physical units, the smallest time step at $z=0$
corresponds to  $1.15\times 10^5$ years.
\section{HALO IDENTIFICATION}

Finding halos in dense environments is a challenge. The most widely
used halo-finding algorithms: the friends-of-friends (hereafter FOF,
e.g., Davis et al.  1985) and the spherical overdensity algorithm (e.g.,
Lacey \& Cole 1994; Klypin 1996) -- are not acceptable (Gelb \&
Bertschinger 1994; Summers et al.  1995). The friends-of-friends (FOF)
algorithm merges together apparently distinct halos if linking radius
is too large or misses some of halos if the radius is too small.  Adaptive
FOF (van Kampen 1995) seems to work better. However, our experiments
show that in practice it is difficult to find an optimal scaling of the
linking radius with the density for a general case.

We have developed a version of the FOF algorithm, which we call
``hierarchical friends-of-friends''. This algorithm uses a fixed
set of hierarchical linking radii and thus does not have problems
adaptive FOF algorithm has.  The algorithms, either adaptive or
hierarchical, cannot work using only geometrical means to identify a
halo.  In the very dense environments they pick up many fake halos.
This can be improved by taking into account dynamical information to
decide whether a halo is real or not. The DENMAX algorithm (Bertschinger \&
Gelb 1991; Gelb \& Bertschinger 1994) or its offspring SKID (Governato
et al.  1997) make a significant progress -- they remove unbound
particles, which is important for halos in groups and clusters. Another
approach to deal with ``flukes'' is to check whether the halos in
question were distinct halos at earlier epochs.  

Recently, Summers et al.  (1995) tried to perfect the idea of Couchman
\& Carlberg (1992) to trace the history of halo merging and to use it
for halo identification. Starting at an early epoch, Summers et al.
identify halos using the FOF algorithm with linking radius
corresponding to the ``virial overdensity'' of 200 and then trace
particles belonging to halos at later times. It appears that it is
impossible to make a working algorithm. Halos interact too violently. A
large fraction of mass is tidally stripped from some halos and a large
fraction of mass is being accreted by others.  However, the idea of
using a set of epochs is too good to be abandoned. To avoid the
problems with the ``direct approach'' (from past to the future), we
have decided to try a reverse logic. Instead of asking the question
``where is now the halo that collapsed at some earlier epoch'', we ask
``did the halo that we find at present exist at an earlier time?''.
Thus, we supplement our hierarchical FOF algorithm with an algorithm
which checks if halos existed at previous moments.

The algorithm which finds halos as maxima of mass inside spheres of a
given overdensity works better than the plain FOF, but no fixed
overdensity limit can find halos in both low and high density
environment.  However, the attractive simplicity of this algorithm
makes it worth in looking for ways of improving it. We will discuss our
improvements to such an algorithm in \S {\SecBDM}.
Besides the problem of finding whether a halo is real, a problem of
``missing'' halos should be kept in mind. As was discussed in the
previous section, while some halos may survive in dense environments, 
others may be destroyed due to numerical effects. These halos will be
missing at $z=0$ and ``present to past'' approach does not help.  

Some of the problems that any halo finding algorithm faces
are not numerical. They exist in the real  Universe. We
select a few of the most typical difficult situations. 

1. {\it A large galaxy with a small satellite.} Examples: the LMC and the
Milky Way or the M51 system.  Assuming that the satellite is bound, do
we have to include the mass of the satellite in the mass of the large
galaxy? If we do, then we count mass of the satellite twice: once
when we find the satellite and then when we find the large
galaxy. This does not seem reasonable. If we do not include the
satellite, then the mass of the large galaxy is underestimated. For
example, the binding energy of a particle at the distance of the
satellite will be wrong. The problem arises when we try to assign
particles to different halos in the effort to find masses of
halos. This is very difficult to do for particles moving between
halos. Even if a particle at some moment has negative energy relative
to one of the halos, it is not guaranteed that it belongs to the
halo. The gravitational potential changes with time, and the particle
may end up falling onto another halo. This is not just a
precaution. This actually was found very often in real halos when we
compared contents of halos at different redshifts. Interacting halos
exchange mass and lose mass. We try to avoid the situation: instead of
assigning mass to halos, we find the maximum circular velocity,
$\sqrt{GM/R}\vert_{max}$, which is a more meaningful quantity than
mass from observational point of view.

2. {\it A satellite of a large galaxy.} The previous situation is now
viewed from a different angle. How can we estimate the mass or the
rotational velocity of the satellite? The formal virial radius of the
satellite is large: it coincides with the virial radius of the host
halo. In order to find the outer radius of the satellite, we analyze
the density profile. At small distances from the center of the
satellite the density steeply declines but then it flattens out and
may even increase. This means that we reached the outer boundary of the
satellite. We use the radius at which the density starts to flatten out
as the first approximation for the radius of the halo. This
approximation can be improved by removing unbound particles and
checking the steepness of the density profile in the outer part.

3. {\it Tidal stripping.} This is not a numerical effect and is not due
to a ``lack of physics''. Very likely this is what happens to real
galaxies in clusters. Their peripheral parts,  responsible for
extended flat rotation curves outside of clusters, are lost when the
galaxies fall into a cluster. Thus, if an algorithm finds that 90\% 
of mass of a halo identified at early epoch is lost, it does not mean that
the halo was destroyed. This is a normal situation. What is left, given
that it still has a large enough mass and radius, is a galaxy halo.

\subsection{Hierarchical friends-of-friends algorithm}

In order to find substructures at vastly different overdensities, we
use hierarchical friends-of-friends algorithm (HFOF).  This algorithm
simply applies the FOF algorithm with a set of different (hierarchical)
linking lengths. In our analysis we use 4 hierarchical levels, in which
case the set consists of linking lengths starting from the small value
$l_{vir}/8$ and larger values obtained by doubling this value: $l
=l_{vir}/4$, $l =l_{vir}/2$, and $l=l_{vir}$.  We call $l =l_{vir}$ the
lowest (with respect to the corresponding overdensity threshold) level
and $l=l_{vir}/8$ the highest level.  The linking length $l_{vir}$
corresponds to the virial overdensity of an object.  We assume the
virial overdensity of 200 and 340 ($l_{vir}=0.2 \bar l$ and
$l_{vir}=0.17 \bar l$) for SCDM/CHDM and for $\Lambda$CDM models,
respectively (e.g., Lahav et al. 1991; Eke et al. 1996); here, $\bar l =
n_0^{-1/3}$ is the mean interparticle separation and $n_0$ is the mean
particle density in the simulation box. For our halos the smallest
linking length, $l=l_{vir}/8$, corresponds to overdensity of $\approx
10^5$.  At each level of the hierarchy identified clusters of particles
(halo candidates) are marked if none of their particles belongs to a
marked higher-level cluster. Finding halo candidates first at the
highest possible level is important because some of higher-level
clusters merge into larger halos at lower levels. In practice, we find
that in high-resolution simulations there is a wealth of small clumps
on all levels in large cluster-size halos, defined on the
lowest level of the hierarchy $l_{vir}$.
 
\begin{figure*}[ht]
\pspicture(0,0)(18.5,17.5)

\rput[tl]{0}(1.,18.){\epsfxsize=17cm
\epsffile{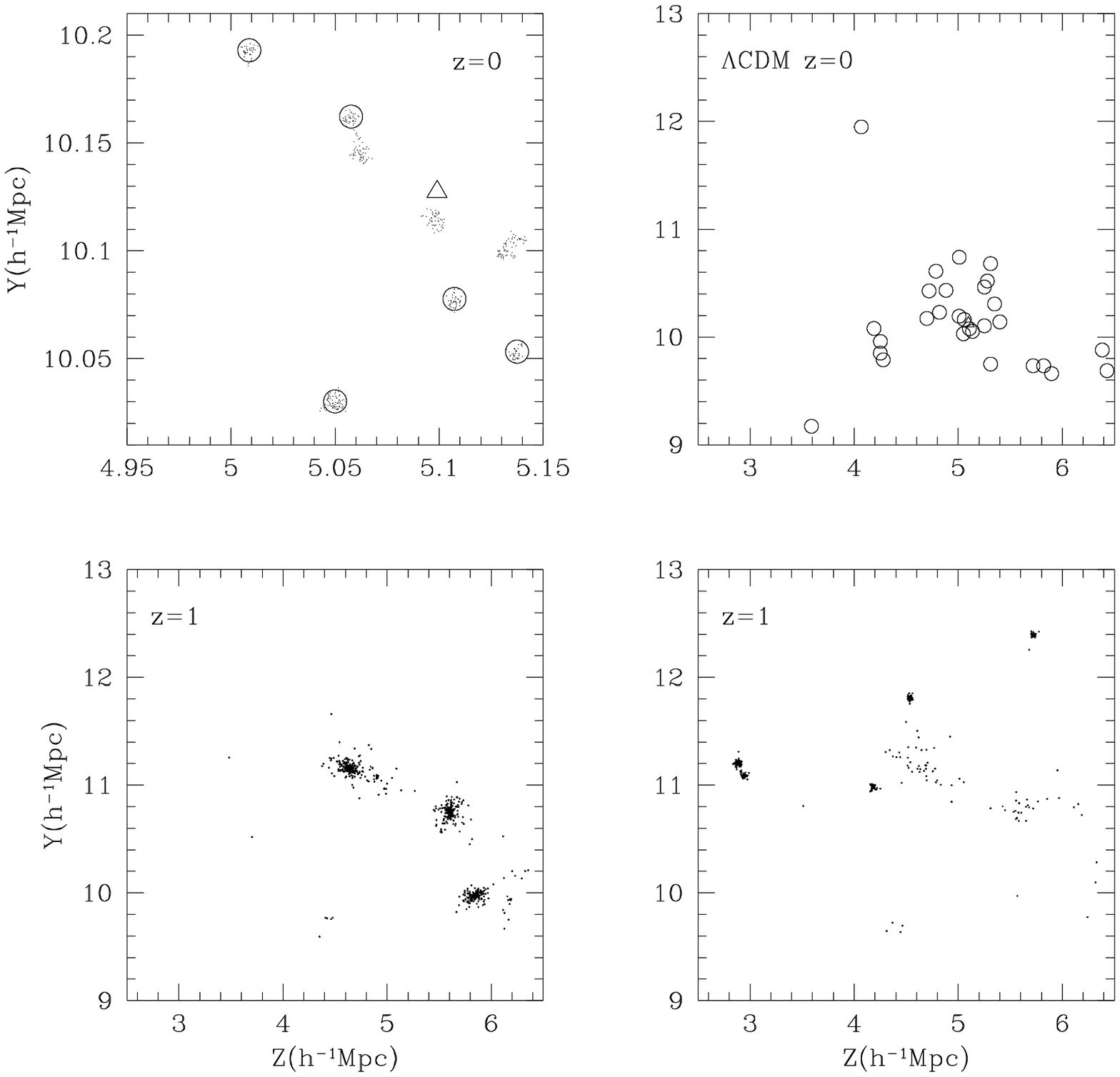}}

\rput[tl]{0}(0.,1.5){
\begin{minipage}{18.5cm}
  \small\parindent=3.5mm {\sc Fig.}~8.--- A halo identification by the
  HFOF.  {\it Top row:} Circles and the triangle represent positions of
  identified halos at the level $l_{vir}/8$ (overdensity of $\approx
  10^5$). Points without circles show fake halos (flukes). Depths of
  the projections are $200\kpch$ for the left panel and $5\Mpch$ for
  the right panel.  {\it Bottom row:} left panel shows position of dark
  matter particles of the fake halos at $z=1$, while the ``real halos''
  are shown in the bottom right panel. Most of particles of the five
  ``real halos'' are in very compact halos whereas the fake halos are
  very puffy (spread over large groups seeing on the top panel of
\end{minipage}
}
\endpspicture
\end{figure*}

Another important feature of this algorithm is use of a particle
distribution at an earlier epoch.  The FOF algorithm works on a
snapshot of the particle distribution and generally tends to identify
particle clusters that are linked at this moment just by chance.
Typical examples would be either a ``bridge'' connecting two particle
clusters or small satellite clusters on highly eccentric orbit that
move temporarily beyond the virial radius of a massive halo. Therefore,
one must check the stability of every identified halo candidate. The
easiest way to make such a check is to find whether a given halo
candidate exists at an earlier moment. We perform such check by running the
HFOF algorithm with the same set of linking lengths using
positions of particles at $z=1$ and check both the existence and
one-to-one correspondence of the progenitor particle clusters. We
consider a candidate halo to be ``stable'' if it has one (or two)
progenitor(s) and it is the only descendant of the progenitor(s).

Details of this analysis are as follows. We select the two most
massive progenitors of each halo and check whether they combined
contain more than a threshold number of particles of the halo at $z=0$. We
search for these progenitors on the next lower level of hierarchy to take into
account the fact that the size of an unevolving object at $z=1$ doubles
in comoving coordinates.  The level is not reduced if it reaches the
virial overdensity. We sum the two most massive progenitors in order to
allow one major merging. In fact, there are very few cases in which 
the masses of three progenitors were of the same order of magnitude.
We used the thresholds 70\%, 50\% and 30\% of mass of the halo at
$z=0$. The threshold 70\% is too high: in many cases halos accrete more
than 30\% of their mass between $z=1$ and $z=0$. The algorithm would
fail to find many halos with this threshold. On the other hand, there
is little difference in the number of identified halos using thresholds of 30\%
and 50\%. However, the algorithm would not find the most massive and
the most dense halos in the center of the large groups because the mass
of those halos increases substantially between $z=1$ and $z=0$ due to
merging with small halos and accretion of single particles. To avoid
this, we also include halos which contain more than a minimum number
(100--500) of particles at $z=1$. We found that the result almost does
not depend on the chosen number because at $z=1$ these progenitors
contain already considerably more particles.

The code checks also whether the particles found in the progenitor
represent a substantial fraction of its mass. The importance of this
criterion is clear from the following example.  Close to the very
massive and very dense halos the FOF algorithm finds
many small clumps. At an earlier moment, all of these small clumps belong
to the same progenitor of the massive cluster around which they were
found at $z=0$. Each lump, however, represents only a tiny fraction of
the progenitor. To avoid misidentifications the algorithm accepts only
halos whose particles represent a substantial fraction of the mass of
that progenitor.  We used values of 10\% and 30\% as thresholds for the
mass fraction of the progenitor . Results are not sensitive to the
particular choice of the threshold. Both criteria (total mass and mass
fraction of the progenitor) result in selection of stable halos.

\begin{figure*}[ht]
\pspicture(0,8.0)(15.0,23.6)
\rput[tl]{0}(-1.25,24.8){\epsfysize=11.5cm
\epsffile{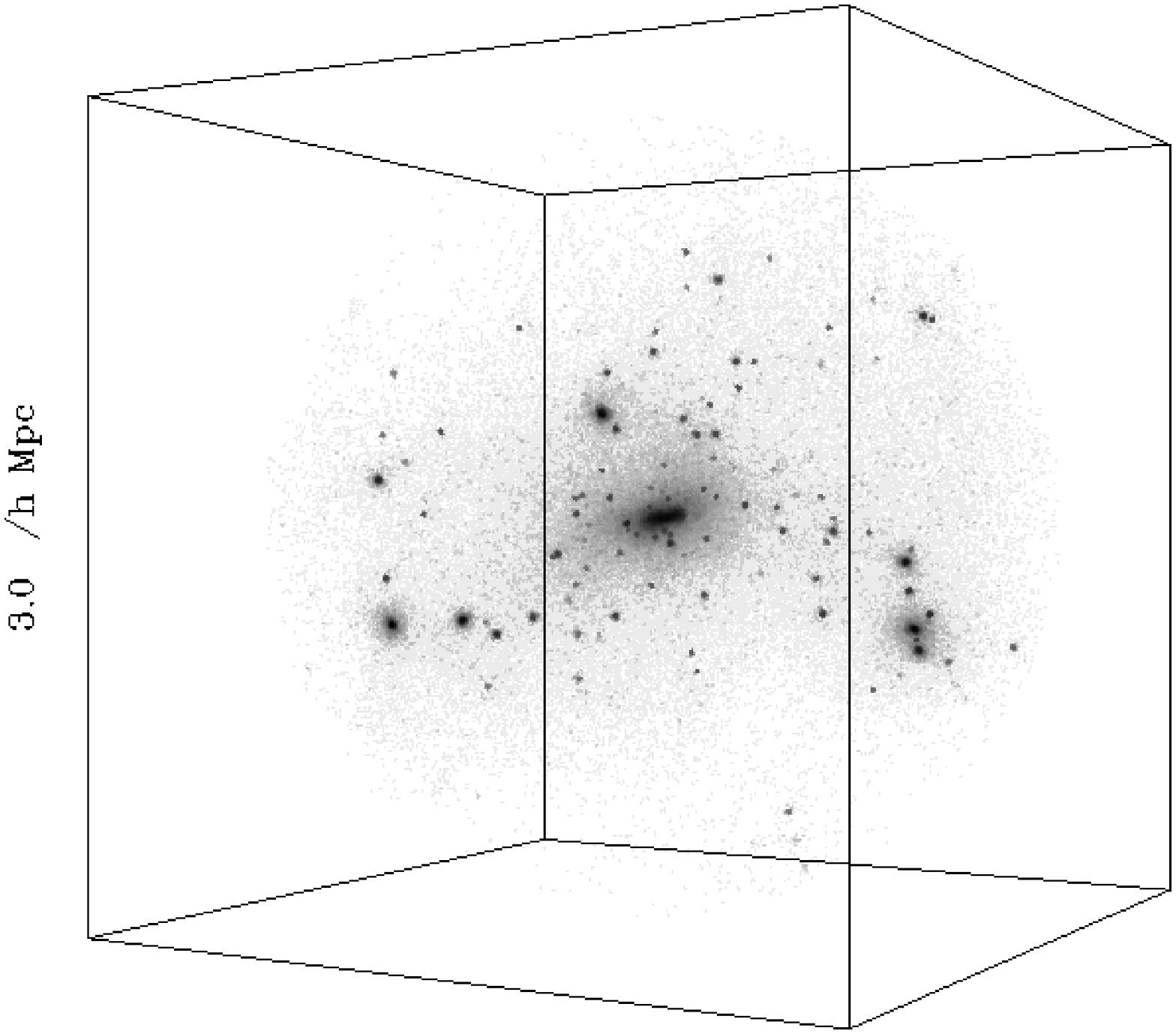}}
\rput[tl]{0}(9.0,24.8){\epsfysize=11.5cm
\epsffile{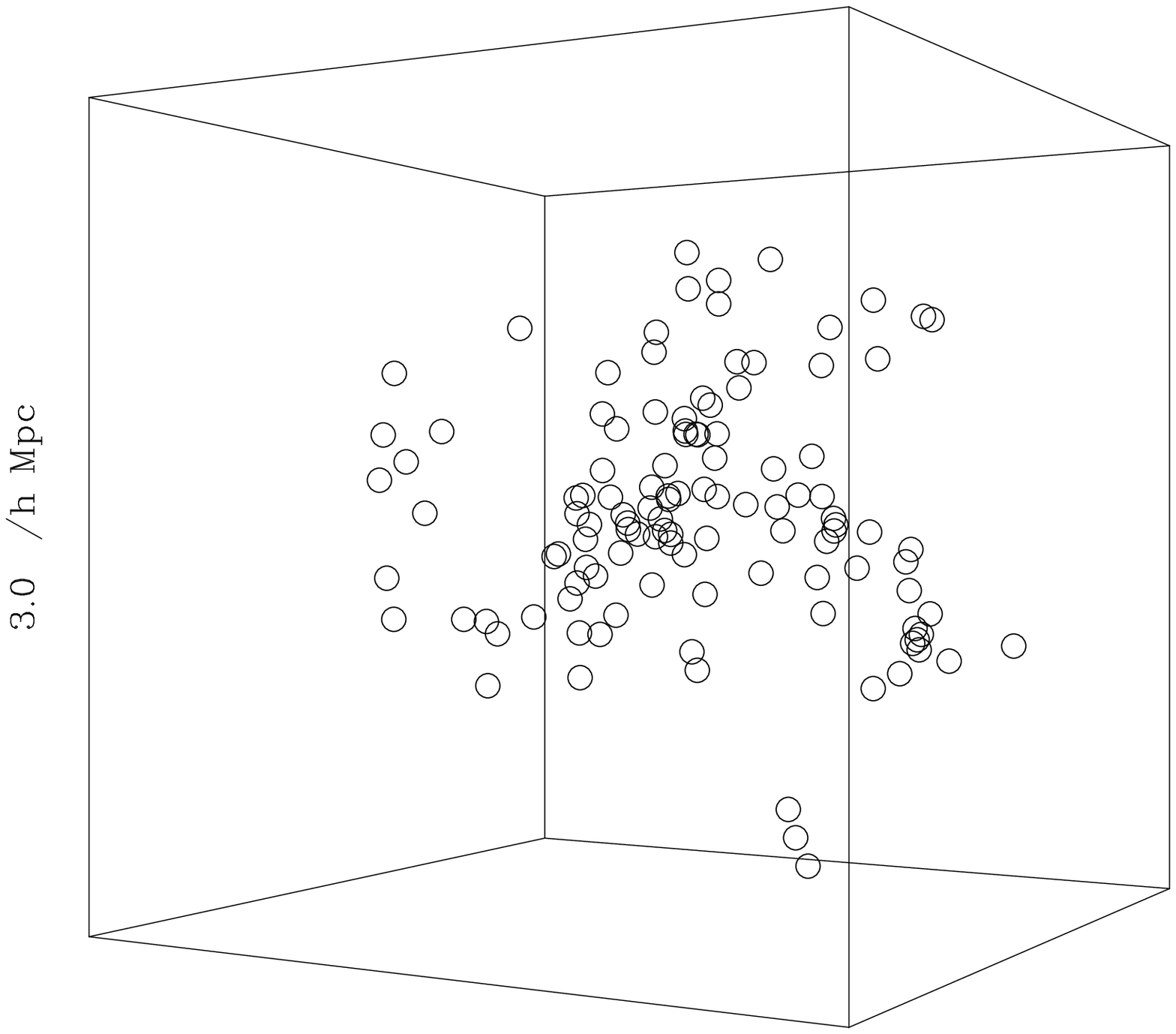}}
\rput[tl]{0}(0.0,11.3){
\begin{minipage}{8.9cm}
  \small\parindent=3.5mm {\sc Fig.}~9.--- An example of a poor cluster
  in a $\Lambda$CDM simulation.  At $z=0$ the cluster has a virial mass
  of $2\times 10^{13}\Mhsun$, virial radius of $500h^{-1} kpc$, and
  velocity dispersion of $\sim 500 km/s$.  The figure shows all DM
  particles ($\sim 250,000$) in a sphere of radius $1.5h^{-1} Mpc$ with
  the centered at the cluster. The particles are color-coded on a
  grey-scale according to the $\log_2$ of the local density (the
  density is estimated as a number of particles at $r<8h^{-1} kpc$ from
  a given particle). Figure {\Halos} shows the halos identified in this
  volume by the bound density maxima algorithm.

\end{minipage}
}
\rput[tl]{0}(9.6,11.3){
\begin{minipage}{8.9cm}
  \small\parindent=3.5mm {\sc Fig.}~10.--- Dark matter halos identified
  by the bound density maxima algorithm in the volume shown in Figure
  {\Groupz} (volume and projection are the same).
\end{minipage}
}
\endpspicture
\end{figure*}

An example of the halo identification in very dense environment is
presented in Figure \HalosGr.  In the top left panel it shows the
inner region of a group-size halo with the triangle denoting the position of
the identified extended massive central ``galaxy'' of the group. The
points surrounded by circles represent dark matter particles in five
identified halos at the level $l_{vir}/8$ (overdensity $\approx
10^5$).   Points
without circles show fake halos (flukes): the halos are found by the
FOF algorithm at this level, but they do not satisfy
 the fraction
of progenitor criterion. Bottom left panel shows position of dark
matter particles of the fake halos at $z=1$, while the ``real halos''
are shown in the bottom right panel. Most of particles of the five
``real halos'' are in very compact halos, whereas the particles
of the fake halos are in extended and ``puffy'' configurations. 

\subsection{Bound density maxima algorithm}

In addition to the hierarchical FOF, we have developed another
halo-finding algorithm, which uses ideas of the DENMAX algorithm
(Bertschinger \& Gelb 1991; Gelb \& Bertschinger 1994). Just as the
DENMAX, our algorithm first finds positions of the density maxima on
some scale and then removes unbound particles inside the halo radius
(hence, the name: Bound Density Maximuma (BDM)). However, the algorithm
finds maxima and removes unbound particles in a different from the
DENMAX way.  The algorithm can work by itself or in conjunction with the
hierarchical FOF. In the latter case, it takes positions of halos from
the HFOF, and then removes unbound particles and finds parameters of
halos.  The version of the BDM code used here is available for use by
astrophysical community (see Klypin \& Holtzman 1997).

In order to find positions of halos we choose a smoothing radius
$r_{sp}$ of a sphere for which we find maxima of mass. This defines the
scale of objects we are looking for, but not exact radii or masses of
halos.  Radius of a halo can be either larger or smaller than $r_{sp}$.
For example, if we are interested in galaxy-size halos, it is
reasonable to choose $r_{sp}\sim (10-15){\ }\kpc$. If we search for
galaxy groups, an appropriate choice is $r_{sp}\sim (200-300)\kpc$.
Then we place a large number of the spheres in the simulation box. The
number of the spheres is typically an order of magnitude or more larger
than the number of expected halos. For each sphere we find its center
of mass and the mass inside $r_{sp}$. The center of the sphere is
displaced to the new center of mass and this process is iterated until
convergence. Depending on specific parameters of the simulations, the
number of iterations ranges from 10 to 100. This process finds local
maxima of mass within $r_{sp}$. Some of the maxima will be found many
times. We remove duplicates and keep only one halo for each maximum.
Halos with too small number of particles (typically 5--10) and halos
with too low central overdensity are removed from the final list.

\begin{figure*}[ht]
\pspicture(0,0)(18.5,19.5)

\rput[tl]{0}(1.,19.5){\epsfxsize=17cm
\epsffile{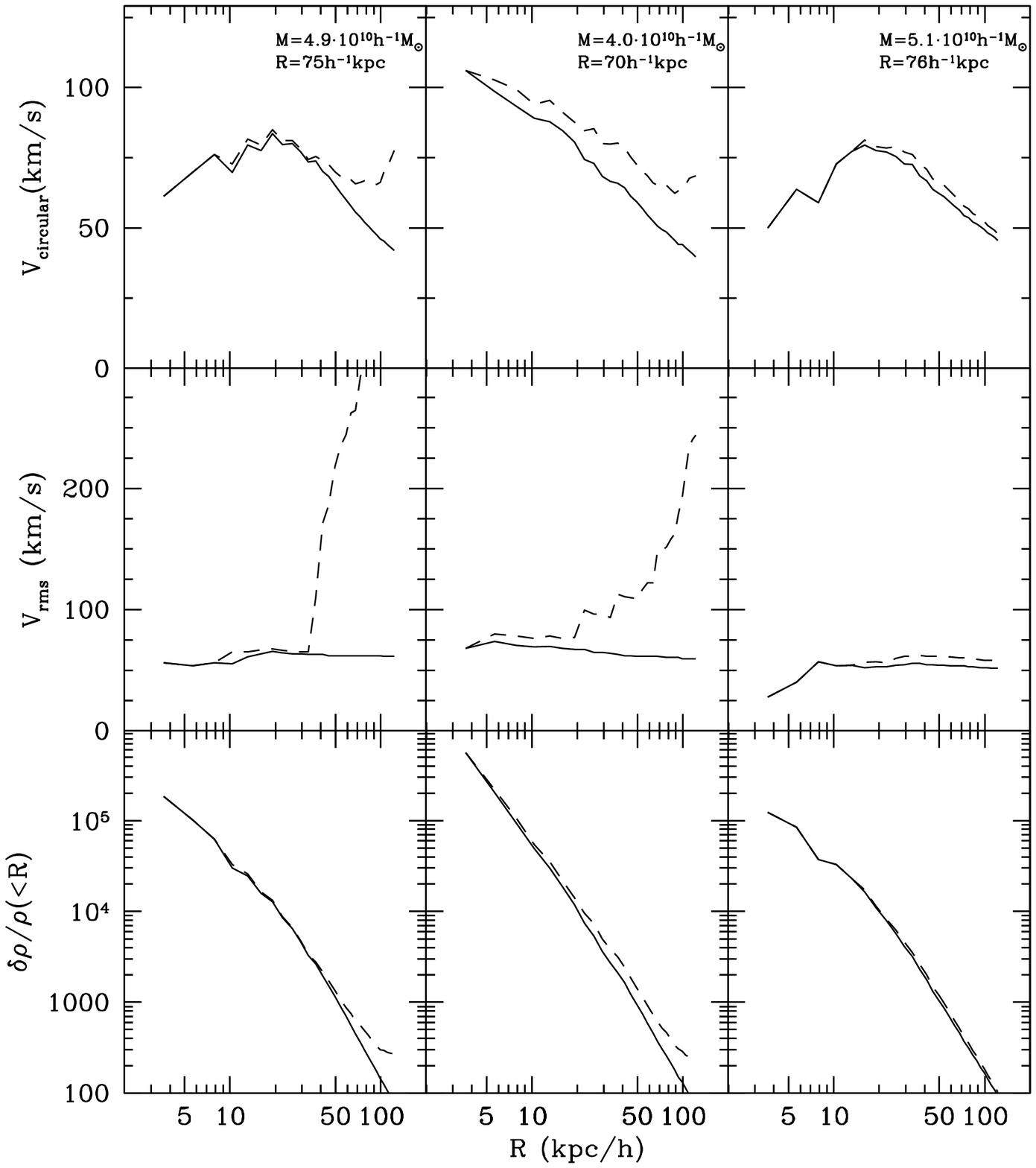}}

\rput[tl]{0}(0.,2.25){
\begin{minipage}{18.5cm}
  \small\parindent=3.5mm {\sc Fig.}~11.--- Examples of profiles of
  small halos in the $\Lambda$CDM model. Each column of plots
  corresponds to the same halo. Mass and radius of each halo (within
  virial overdensity 340) are shown in the top panels.  The dashed
  curves are for the halos before the removal of unbound particles, and
  the full curves are for bound particles only.  The top row of panels
  shows the circular velocity $V_{circular}=(GM(R)/R)^{1/2}$. The
  middle row shows velocity dispersion of the dark matter particles,
  and the bottom row presents the overdensity profiles. The right halo
  is an example of an isolated halo in which most particles are bound
  to the halo. The two halos on the left show examples of small
  satellite close to a large halo. In the central $\sim 20\kpch$ part
  of both halos the velocity dispersions are small and almost constant.
  But at $30\kpch$ the velocity dispersion starts a dramatic increase
  indicating presence of a massive object.
\end{minipage}
}
\endpspicture
\end{figure*}

\begin{figure*}[ht]
\pspicture(0,0)(18.5,18.)

\rput[tl]{0}(1.,18.5){\epsfxsize=17cm
\epsffile{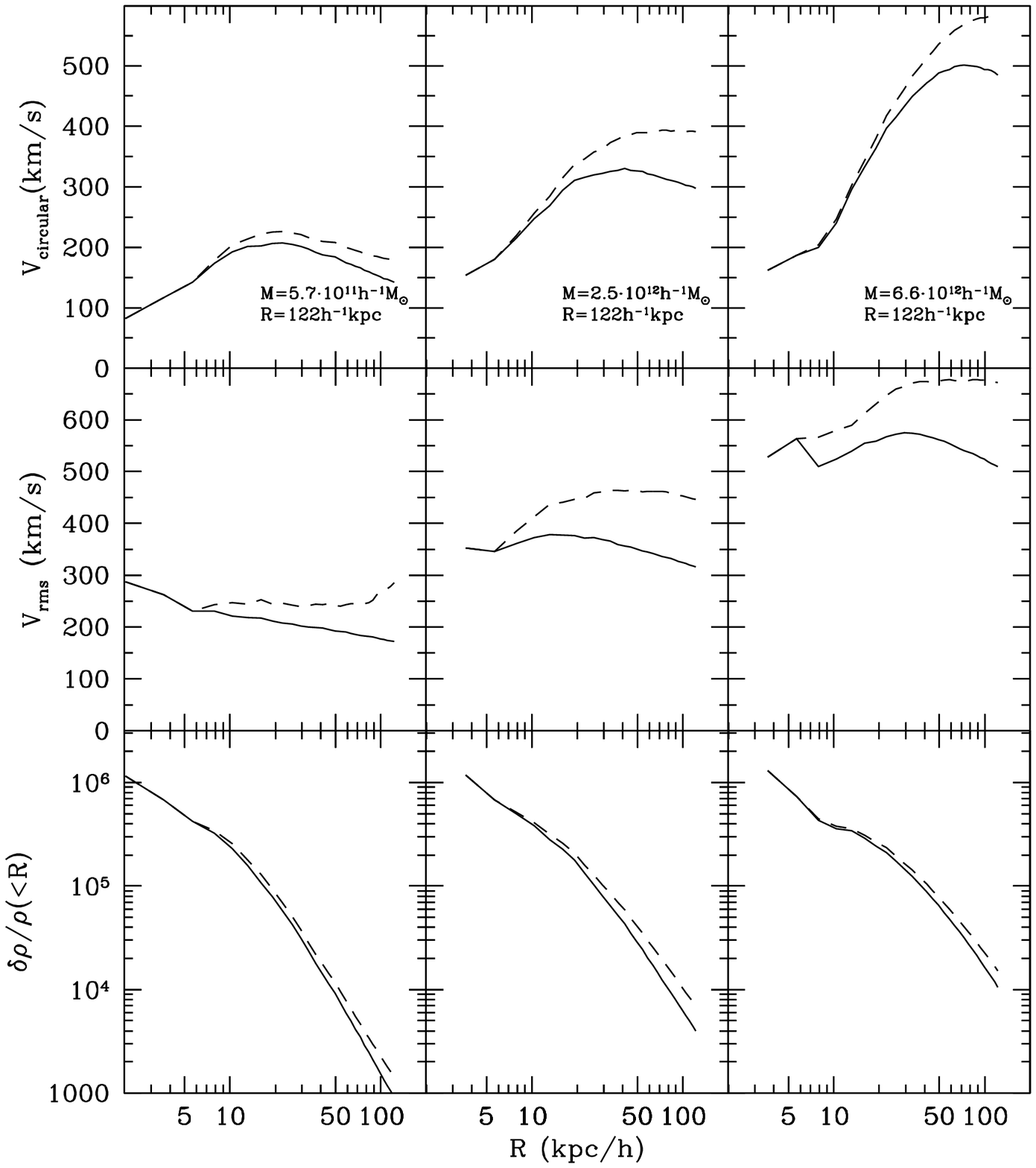}}
\rput[tl]{0}(0.,1.){
\begin{minipage}{18.5cm}
  \small\parindent=3.5mm {\sc Fig.}~12.--- The same as in Figure
  {\Profsm} but for medium and large halos. In this case the masses are
  given for constant outer radius of $122\kpch$. The mean overdensity
  at that radius for all halos is significantly larger than the virial
  value. But the maximum of rotational velocity is well within the
  distance.
\end{minipage}
}
\endpspicture
\end{figure*}

Figures {\Groupz} and {\Halos} illustrate typical
output of the algorithm. Figure {\Groupz} shows a group-size halo 
identified in our $30\hMpc$ $\Lambda$CDM simulation. The particles,
shown in the figure inside the sphere of radius $1.5\Mpch$ centered on
the group, are color-coded on a grey-scale according 
to the logarithm of the local density, estimated as a number of
particles within $8\hkpc$ of the particle. Figure {\Halos} shows dark
matter halos identified by the BDM algorithm in this volume. All of the
halos, that can be seen in Figure {\Groupz} as tight dark clumps of
particles, are identified by the halo finder.

Once centers of potential halos are found, we start the procedure of
removing unbound particles and finding the structure of halos.  We
place concentric spherical shells around each center. For each shell we
find mass of the dark matter particles, mean velocity, and the velocity
dispersion relative to the mean. In order to determine whether a
particle is bound or not, we estimate the escape velocity (Eqs.\EqVmax)
at the position of the particle. If velocity of a particles is larger
than the escape velocity, it is assumed to be unbound. We estimate the
maximum rotational velocity $V_{max}$ and radius of the maximum
$r_{max}=2r_s$ using the density profile for the halo. Because
$V_{max}$ and $r_{max}$ must be found before the unbound particles are
removed and because the mean velocity is also found using all particles
(bound and unbound), the whole procedure can not be done in one step.
We start by artificially increasing the value of the escape velocity by
a factor of three.  Only particles above the limit are removed. We find
new density profile, new mean velocities, new $V_{max}$ and $r_{max}$.
The escape velocity is again increased, but this time by a smaller
factor. The procedure is repeated 6 times. The last iteration does not
have any extra factors for the escape velocity.

Removal of unbound particles is crucial
in the case when a halo with a small internal velocity dispersion moves
inside a large group. For example, if a halo with a circular velocity
of $100\kms$ moves  with velocity $500\kms$ inside a group, a dark matter
particle bound to the group has kinetic energy 25 times larger than the
kinetic energy of a particle of the halo. Even if only 1/10th of
particles within the halo radius belong to the group, the whole halo
will have positive energy and will be treated as fake. Removal of
unbound particles salvages the halo even if the real halo particles
constitute as little as 1/4 of the total number of particles within the
halo radius. This estimate is valid in the case of a compact halo
moving through homogeneous field of high velocity particles of the
group. The situation is worse if the particles of the high velocity
field are very lumpy. The worst case is the collision of two equal mass
halos. At the moment when the halos overlap, the code does not find any
bound component -- both halos are missed. The chance of such event is
very small because the distance between centers of halos should be
smaller than $\sim (10-15)\hkpc$.

The effect of the unbound particle removal on the halo profiles is
illustrated in Figures {\Profsm} and \Profmed. The halos were
identified in the simulation of the $30\Mpch$ $\Lambda$CDM simulation.
In the Figure {\Profsm} the right column shows circular velocity,
velocity dispersion, and density profiles of a ``normal'' halo with a
small fraction of unbound particles, primarily in the outer regions.
The middle and left columns show profiles of small satellite halos
located inside or close to a massive halo. In the central $\sim
20\kpch$ of both halos the velocity dispersions are small and almost
constant. The circular velocities are about what one should expect for
these values of the velocity dispersions. But at $30\kpch$ the velocity
dispersion increases dramatically, indicating a presence of a dense
background of fast moving particles belonging to the massive halo.
Figure {\Profmed} shows typical examples of medium- and high-mass
halos.

\subsection{Comparison of the HFOF and BDM algorithms }

The goal of both of the described algorithms is to find positions of
stable halos in a given simulation. Nevertheless, the two algorithms
are quite different. The HFOF, for example, computes mass, spin
parameter, angular momentum, shape and total binding energy of the
halos without removal of unbound particles. The algorithm, however,
does not assume spherical symmetry in these calculations. The BDM
algorithm computes properties of the halos (e.g., mass, velocity,
density profile) after removing unbound particles, the procedure based
on the assumption of spherical symmetry. 
Therefore, one must expect that the algorithms may compute slightly
different properties for the same halos.

Nevertheless, the most important information determined by halo finders
is positions of DM halos.  Our tests show that the algorithms find
exactly the same halos, if the halos contain more than a couple hundred
particles. The agreement is about 95 \% for halos with more than 50
particles and about 90 \% for halos with more than 30 particles. The
small differences in the threshold criteria for the selecting halos
account for the 10 \% differences at the low mass end of the halo
distribution. But overall, the agreement is very good. For example, the
five small and the big central halo in a high density region shown on
the top left panel of Figure {\HalosGr} have been found by both
algorithms. We also compare halos in a statistical way.  In Figure
{\CorrFun} we compare the correlation functions of halos found by the
two algorithms. For this comparison we have selected halos with rather
low limit in maximum circular velocity $V_{max} > 90$ km/s. In case of
the $\Lambda$CDM simulation (lower panel) the two correlation functions
coincide within the expected scatter due to the statistical noise. This
indicates that the thresholds are chosen in an equivalent manner. In
case of the CHDM simulation (upper panel) the correlation functions
differ more. Very likely this happened because of the slight mismatch
in the mass limits. As expected, the two correlation functions differ
mainly on scales less than 100 kpc due to different thresholds used in
the algorithms.

%
\section{RESULTS}
%
\subsection{Luminosity function of galaxies and M/L in groups}
The observed tight correlation between the 21-cm line width $W\approx
2V_{circ}$ and infrared luminosities for spiral galaxies (e.g.,
Aaronson \& Mould 1983; Bureau, Mould, \& Staveley-Smith 1996; Willick
et al.  1996; Giovanelli et al.  1997) can be used to estimate
luminosities of galaxy-size halos in $N$-body simulations (an
alternative method for assigning luminosities to the DM halos and
constructing the luminosity function was proposed recently by Roukema
et al. 1997). This is probably the best one can do when dealing with
dissipationless simulations.  Unfortunately, the observed Tully-Fisher
relation is not defined as accurately as one would hope for. This is
especially true for low-luminosity galaxies. In the following, we apply
the empirical Tully-Fisher relation determined for galaxies with
absolute blue magnitudes $m_B\leq-15$. 
{\pspicture(0.5,-1.5)(13.0,10.5)
\rput[tl]{0}(-0.5,12.){\epsfxsize=10.5cm
\epsffile{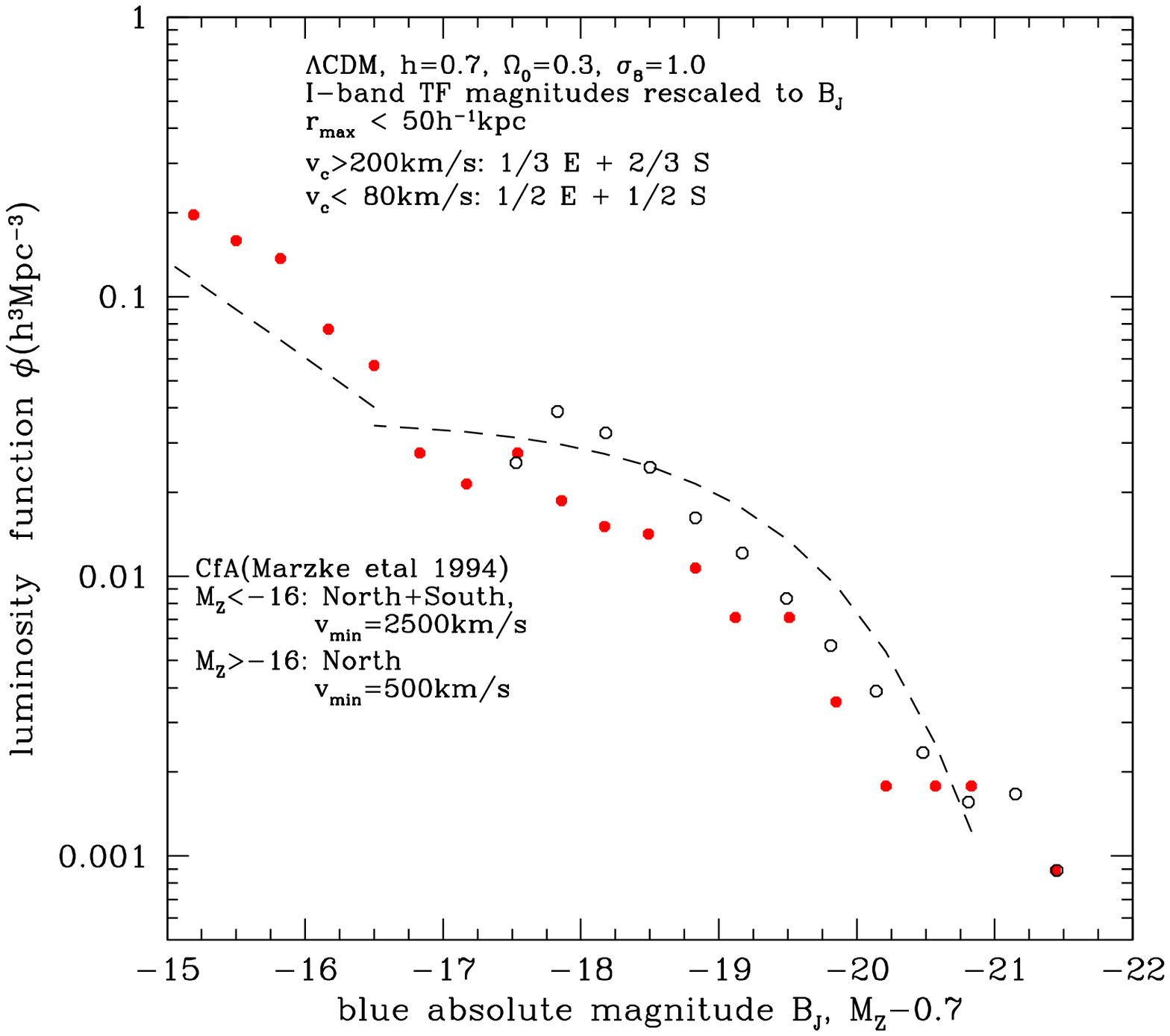}}
\rput[tl]{0}(0.5,1.5){
\begin{minipage}{8.7cm}
  \small\parindent=3.5mm {\sc Fig.}~13.--- Comparison of the luminosity
  function of galaxy-size halos in the LCDM simulations with the CfA
  data (dashed curve). The solid circles show results for the $15\Mpch$
  box. The results for $30\Mpch$ box are shown with open circles. The
  first bin of the $30\Mpch$ simulation at $M=-21.45$ has 8 galaxies;
  the bin at $M=-17.8$ has 350 galaxies.  There are 146 galaxies
  brighter than $B_J=-20$ and 1340 brighter than $B_J=-18$ in the
  $30\Mpch$ box simulation.
\end{minipage}
}
\endpspicture}
It should be kept in mind,
however, that there may be significant systematic deviations from the
relation used here for galaxies with magnitudes of $m_B > -17$.

Elliptical galaxies pose another problem. The Faber-Jackson relation
indicates that velocity dispersion (and thus the dark matter mass) can
be used to estimate galaxy luminosity. The relation, however, is not
very tight. In principle, by tracing the merging history of each halo,
we can make more realistic estimates of the star formation rates and
the luminosities. Here, we will use a simple but reasonable
prescription; we assume that an elliptical galaxy is $\sim 1$ magnitude
dimmer than a spiral galaxy with the same maximum circular velocity.
This assumption is motivated by the fact that mass-to-light ratio of
elliptical is 2.5-3 times higher ratio than that of sprials spirals.
It is likely that the fraction of ellipticals is significant only at
the the high and low mass ends of mass function. We assume therefore
that all halos with $V_{circ}>350\kms$ and half of the halos with
$V_{circ}<80\kms$ host elliptical galaxies.  Halos in the ``grey area''
$V_{circ}=(200-350)\kms$ have a gradually increasing probability to
host an an elliptical: $\propto 1/3(V/200 km/s)^2$.

It is important to stress the use of maximum circular velocity and/or
velocity dispersion of halo as a mass indicator. Problems of finding
dark matter halos and determining their masses in numerical simulations
highlighted in \S 4, make it virtually impossible to make a meaningful
assignment of mass to a halo in a group or a cluster. On the other
hand, velocity dispersion or maximum circular velocity can be
determined more or less reliably with even moderate particle
statistics. 
{\pspicture(0.5,-1.5)(13.0,8.)
\rput[tl]{0}(-0.5,10.){\epsfxsize=10.5cm
\epsffile{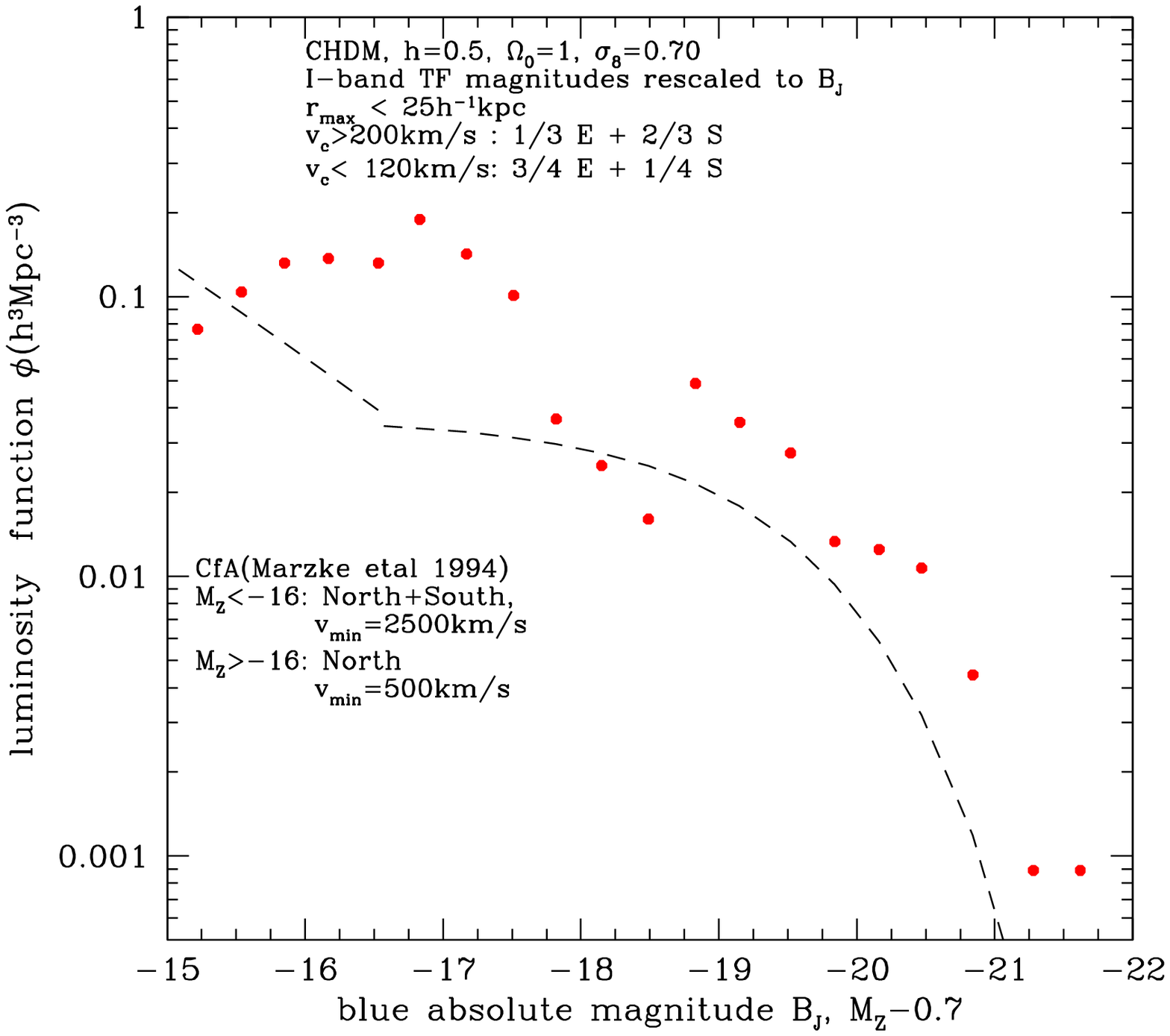}}
\rput[tl]{0}(0.5,-0.5){
\begin{minipage}{8.7cm}
  \small\parindent=3.5mm {\sc Fig.}~14.--- The same as in Figure
  \LumLCDM, but for the CHDM $15\Mpch$ box simulation.
\end{minipage}
}
\endpspicture}
We believe therefore that use of the maximum circular
velocity as a proxy for halo mass is more attractive than use of
potentially unreliable and biased mass estimates. 

In order to estimate luminosities of ``galaxies'' hosted by DM halos
using their maximum circular velocities, we use the following
Tully-Fisher relation in the I-band:
$M_I-5log(h)=-21.0-6.8(\log{W}-2.5)$. The slope of the relation is as
given from Willick et al.  (1996) for field spirals, while the zero
point was adopted from Giovanelli et al.  (1997). The I-band magnitudes
were shifted to the blue magnitudes as $M_B=M_I+1.5$ (Pierce \& Tully,
1992).  Because circular velocities in observations are never measured
at large galactocentric distances, we set an upper limit of $r_{max} <
50\kpch$ for the radius of the maximum of the rotational velocity.

The main caveat of the above luminosity assignment scheme is that we
use the maximum circular velocities of halos, discarding the disk
contribution. However, as we have discussed in \S 2.3, this may result
in maximum error of $30\%$ in the circular velocity.  Also, Figure
{\DVevol} shows that maximum circular velocity of halos orbiting in
clusters may decrease by $\sim 20-50\%$ due to the tidal stripping. The
luminosity assignment may be somewhat biased for such halos if baryons are not
stripped as efficiently as the dark matter. 

Figure {\LumLCDM} shows the luminosity function (LF) of galaxy-size
halos identified in the two of our $\Lambda$CDM simulations. The solid
circles show results for the $15\Mpch$ box. There is a significant tail
of low luminosity galaxies ($B_J > -16.5$), which matches the
corresponding tail of the CfA luminosity function (Marzke et al.  1994)
and the luminosity function of cluster galaxies (Smith et al.  1997;
Lopez-Cruz et al. 1997). The simulation of the larger box of $30\Mpch$
(open circles) has insufficient mass resolution to probe this low-mass
tail, but the two LFs are consistent in the region of overlap.  We
believe that the LFs in Figure {\LumLCDM} are most reliable in the
magnitude range of $B_J= -17-20$. In this range the number of of halos
in each luminosity bin is sufficiently high to make the poisson errors
insignificant and results do not depend on the assumed fraction of
ellipticals or on the maximum allowed radius for the circular velocity.
The first bin of the $30\Mpch$ simulation at $M=-21.45$ contains 8
halos, while the bin at $M=-17.8$ contains 350 halos. The luminosity
function in the $\Lambda$CDM model in this range of magnitudes is
systematically lower by a factor of 1.5--2 than the CfA luminosity
function. This is consistent with deeper samples, which give lower
normalization for the luminosity function (e.g., Loveday et al.  1992).

The luminosity function of galaxy-size halos in the CHDM simulation is
significantly higher than both the $\Lambda$CDM and the CfA luminosity
functions.  With the same set of parameters as for the $\Lambda$CDM
model, the LF in the CHDM model is a factor of 4-5 higher than the
luminosity function in the CfA catalog. In order to reconcile the model
with the observational data, we reduced the limit on the radius for the
rotational velocity to $r_{max}=25\kpch$ and raised the fraction and
the limiting magnitude for small elliptical galaxies. Figure {\LumCHDM}
shows the CHDM luminosity function that best matches the CfA luminosity
function. It is still systematically higher than the CfA LF,
but it might be acceptable due to small volume of the simulation.
\subsection{Halo dynamics in groups: velocity bias, $M/L$, and
  constrains on $\Omega$}

The observed mass-to-light ratios of galaxy groups and clusters,
$(M/L)_B\approx 150-400$ (e.g., Bahcall, Lubin \& Dorman 1995), are
often used as an argument in favor of the low-density universe with
$\Omega_0\approx 0.2-0.3$. We have used the halos identified in poor
galaxy clusters and groups in our simulations to study their dynamical
properties and estimate the mass-to-light ratio of these clusters using
the same prescription to assign luminosity to halos as was used in the
previous section.  
Figures {\MassVelocityLCDM} and {\MassVelocityCHDM}
present different properties of groups of galaxies in the simulations.
Centers of the groups were found using search radius of
$r_{sp}=0.250h^{-1}$ Mpc and no removal of unbound particles was done
in this case.  Radius of groups was estimated at the overdensity limit
of 200 for the CHDM model and of 340 for the $\Lambda$CDM model.  The
results clearly indicate that the M/L ratio increases with the mass of
the group. However, there is an indication that in the $\Lambda$CDM
simulation $M/L$ ratio for massive groups flattens at the level of
$\approx 300h^{-1}(M_{\odot}/L_{\odot})$.  Note, that models with
$\Omega_0=0.3$ ($\Lambda$CDM) and $\Omega_0=1$ (CHDM) reproduce the
observed $M/L$ ratios equally well, although due to the small volume,
we can only probe masses of $\lesssim 3\times 10^{13}h^{-1}{\ 
  }{M_{\odot}}$ in the CHDM simulation.  It appears thus that
mass-to-light ratio of galaxy groups of mass $\lesssim 3\times
10^{13}h^{-1} {\rm M_{\odot}}$ is not a very good indicator of the
total matter density in the universe.

Several authors have suggested a possible existence of the velocity
bias $b_{v}\equiv \sigma_{halo}/\sigma_{dm}$: systematic difference
between rms velocities of galaxies and dark matter particles (Carlberg,
Couchman \& Thomas 1990; Carlberg 1994; Colafrancesco,
Antonuccio-Delogu \& Del Popolo 1995). The existence of the velocity
bias would have impact on the determination of cluster masses using
galaxy dynamics and other analyses.  
{\pspicture(0.5,-1.5)(13.0,14.)
\rput[tl]{0}(-0.5,14.){\epsfxsize=10.5cm
\epsffile{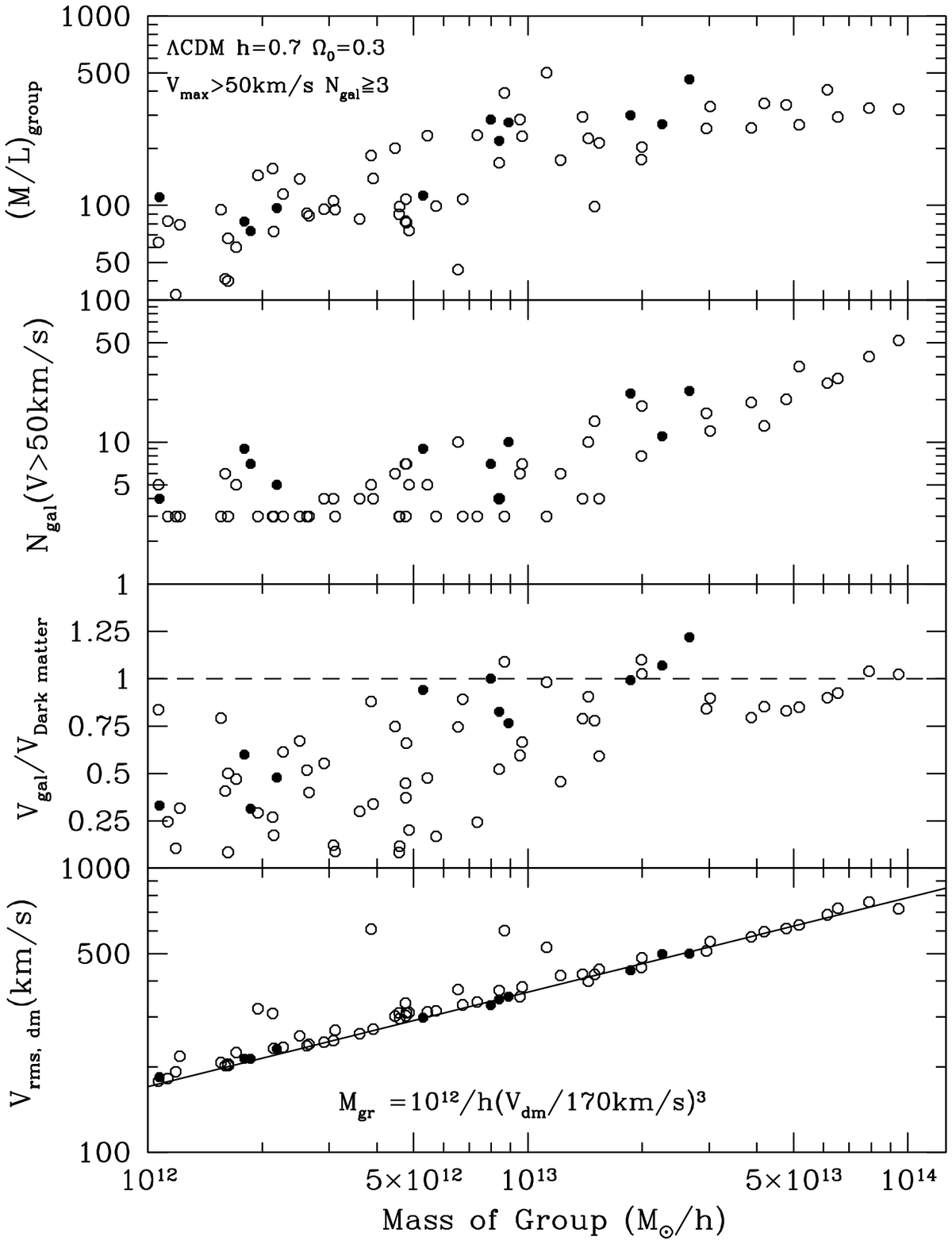}}
\rput[tl]{0}(0.5,3.25){
\begin{minipage}{8.7cm}
  \small\parindent=3.5mm {\sc Fig.}~15.---  Properties of groups in the
 $\Lambda$CDM simulations as a function of the total mass of the
 group. $V_{rms, dm}$ is the root-mean-square velocity of dark matter
 particles in the group. The ratio of the rms velocities of galaxies in
 the group to $V_{rms, dm}$ is shown in the second panel. Results for
 groups with more than 3 satellites are shown. In small groups with mass
 less than $10^{13}\Mhsun$ there is on average a significant velocity
 bias, which is likely related to the dynamical friction. There is no
 indication of the velocity bias for larger groups. The number of
 galaxies with $V_{circ} > 50\kms$ in a group is shown in the third
 panel from the bottom. The mass-to-light ratio of groups (in units of
 $h^{-1}\Msun/L_{\odot,B}$) is shown in the top panel. 
\end{minipage}
}
\endpspicture}
Figures {\MassVelocityLCDM} and
{\MassVelocityCHDM} show that there is no evidence for strong velocity
bias for halos in groups in our simulations; on average, for objects
$\gtrsim 10^{13}h^{-1} M_{\odot}$ the rms velocities of galaxies and
dark matter particles are equal.  Although theoretical arguments
suggest that a mild velocity bias should exist (due, for example, to
effects of dynamical friction), the uncertainties of current
simulations favor absence of the velocity bias ($b_{v}=1$) and excludes
values of the bias $b_{v}\lesssim 0.8$, allowing, however, for a
possible existence of a mild bias $b_{v}\approx 0.8-0.9$. Note that for
small groups of galaxies ($10^{12}h^{-1} M_{\odot}\lesssim
M_{vir}\lesssim 10^{13}h^{-1} M_{\odot}$) a significant velocity bias
is observed. This bias could probably be attributed to the strong
dynamical friction effects operating in these systems.

\subsection{Small scale two-point halo correlation function}
Analyses of recently completed galaxy surveys resulted in a very
accurate determination of the galaxy two-point correlation function,
$\xi(r)$, at the scales of $\gtrsim 20-50\hkpc$ (e.g., Baugh 1996).
The comparison of the observed $\xi(r)$ with the two-point correlation
function of mass in dissipationless $N$-body simulations has shown that
an antibias of galaxies is required at small scales in order to
reconcile the 
{\pspicture(0.5,-1.5)(13.0,10.)
\rput[tl]{0}(-0.5,10.){\epsfxsize=10.5cm
\epsffile{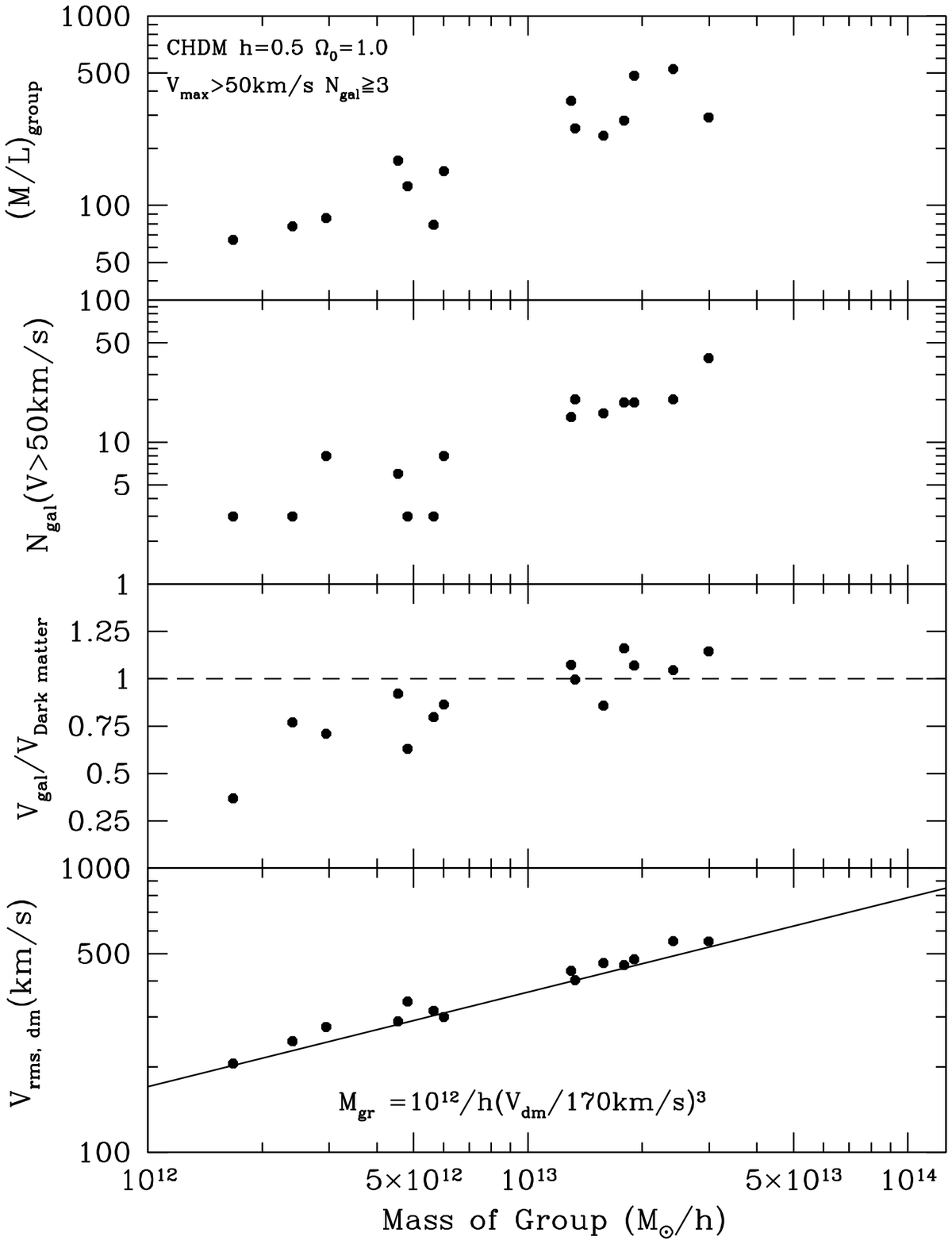}}
\rput[tl]{0}(0.5,-1){
\begin{minipage}{8.7cm}
  \small\parindent=3.5mm {\sc Fig.}~16.---   Properties of groups in the CHDM 
simulation.
\end{minipage}
}
\endpspicture}
models with observations (Klypin, Primack \& Holtzman
1996; Jenkins et al. 1998).  In order to check whether any significant
antibias exists for the dark matter halos (which could be associated
with observed galaxies), we have constructed the halo-halo two-point
correlation function for the halos found in our simulations using both
hierarchical friend-of-friend and bound density maxima algorithms.
Figures {\CorrFun} and {\CorrMass} show correlation functions of halos
and dark matter in the simulations. The correlation function is clearly
affected at scales $\gtrsim 500\hkpc$ by the finite box size (as shown
by comparison of $\xi(r)$ for small and large box simulations of the
$\Lambda$CDM model). However, it is interesting to examine the relative
behavior of the dark matter and halo correlation functions at scales
$\lesssim 500\hkpc$.  At very small scales (less than $100\kpch$) the
galaxies are more clustered (biased) relative to the dark matter; at
larger scales the effect is the opposite: galaxies are slightly
antibiased. This is observed in all simulations and it is valid for all
limits on masses of galaxies.  It is interesting that the antibias of
the magnitude 0.7--0.9 seen in the Figure {\CorrFun} is almost exactly
what is needed for the $\Lambda$CDM model to be compatible with
observational data on the power spectrum in the range of wavenumbers
$k=(0.1-1)h\Mpc^{-1}$ (Klypin et al.  1996; Smith et al.  1998; Jenkins
et al. 1998).  The antibias of halos at these small scales is very
likely related to the tidal destruction and dynamical friction of halos
in groups.  The $100\kpch -1\Mpch$ range is the range where we expect
both of these processes to work. This conjecture seems to be confirmed
by results of larger simulations (Col\'{\i}n et al. 1998; Kravtsov \&
Klypin 1999).
{\pspicture(0.5,-1.5)(13.0,12.5)
\rput[tl]{0}(-0.75,12.5){\epsfxsize=10.5cm
\epsffile{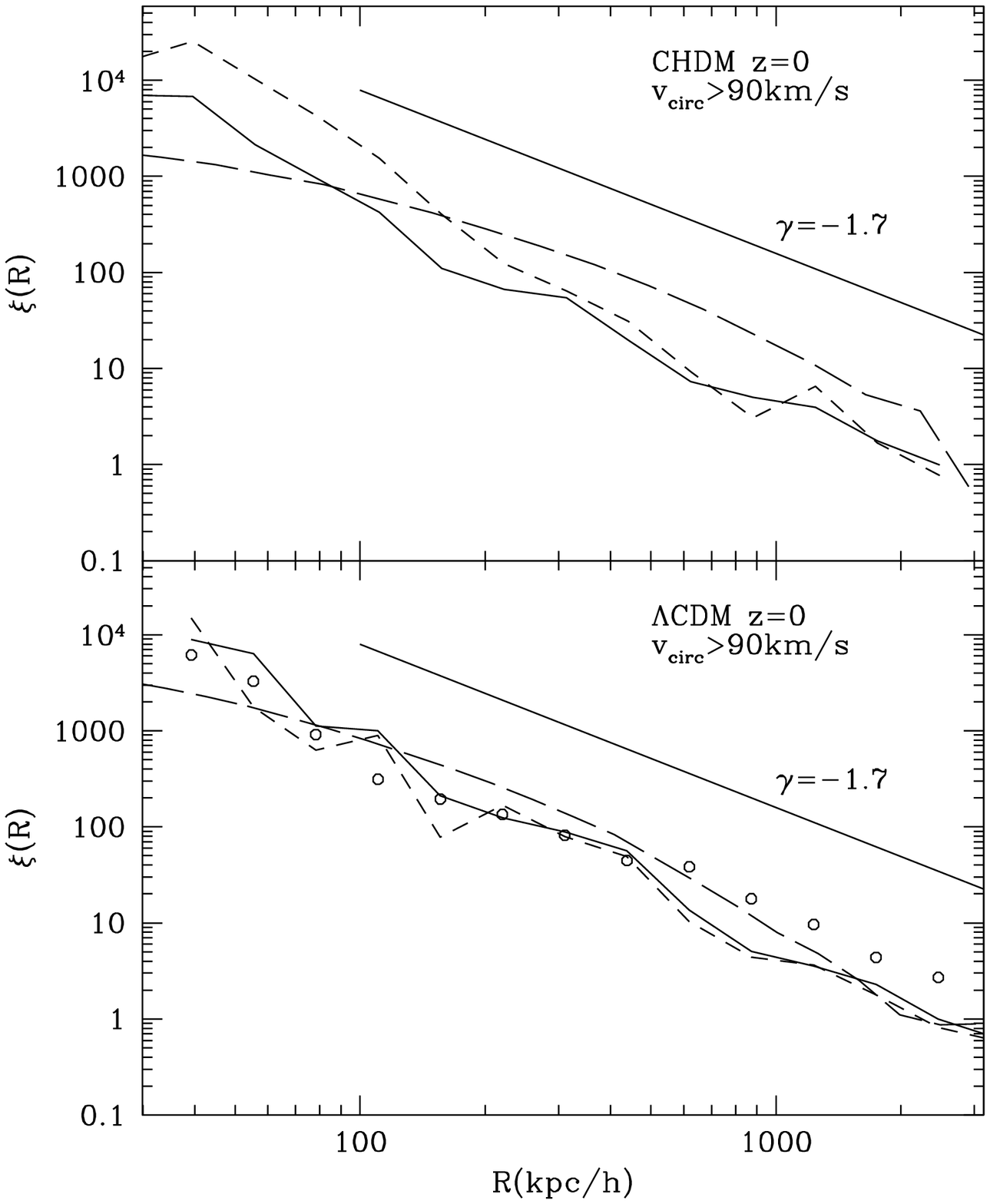}}
\rput[tl]{0}(0.5,2){
\begin{minipage}{8.7cm}
  \small\parindent=3.5mm {\sc Fig.}~17.--- Correlation functions of the
  dark matter (long dashed curves) and galaxy-size halos in the
  $\Lambda$CDM (bottom) and the CHDM simulations (top) in the $15\Mpch$
  boxes. Galaxies with rotational velocity larger than $90\kms$ were
  selected. Short-dashed curves are for galaxies identified with the
  hierarchical friends-of-friends algorithm. The solid curves are for
  halos identified using bound density maxima algorithm. The circles
  are for halos identified using the bound density maxima algorithm in
  the $30\Mpch$ box simulation with the same limit on the rotational
  velocity.
\end{minipage}
}
\endpspicture}
%
\section{CONCLUSIONS} 
%
We have presented arguments that the overmerging problem is mostly due
to inability of a numerical code to provide a sufficient numerical
resolution to prevent tidal destruction of galaxy-size halos by the
tidal forces of group or cluster and have estimated the resolution needed to
prevent such disruption. We argue that although
energy dissipation by the baryonic component helps galaxies to
survive in clusters, at distances $\gtrsim (50-70)\kpch$ from
the cluster center the gravity of the dark matter alone is enough to
keep them alive.

The main result of this work is estimate of the numerical resolution
needed to overcome the overmerging problem.  The results of our
analytic estimates and numerical experiments show that although it is
feasible to overcome overmerging in pure $N$-body simulations, resolution
required to avoid artificial destruction of galaxy-size halos (mass
$\gtrsim 10^{11}h^{-1} {\rm M_{\odot}}$) is quite high. For viable CDM
models and realistic halo profiles this resolution is $\lesssim 2h^{-1}
{\rm kpc}$ in force and $\lesssim 10^{9}h^{-1} {\rm M_{\odot}}$ in
mass. This requires simulations of $>10^7$ particles with dynamic range
of $10^5$ in spatial resolution for statistically significant
cosmological volumes ($\sim 100 {\rm Mpc}$), which remains challenging
with the current computers and numerical codes. 
{\pspicture(0.5,-1.5)(13.0,12.)
\rput[tl]{0}(-0.75,12.){\epsfxsize=10.5cm
\epsffile{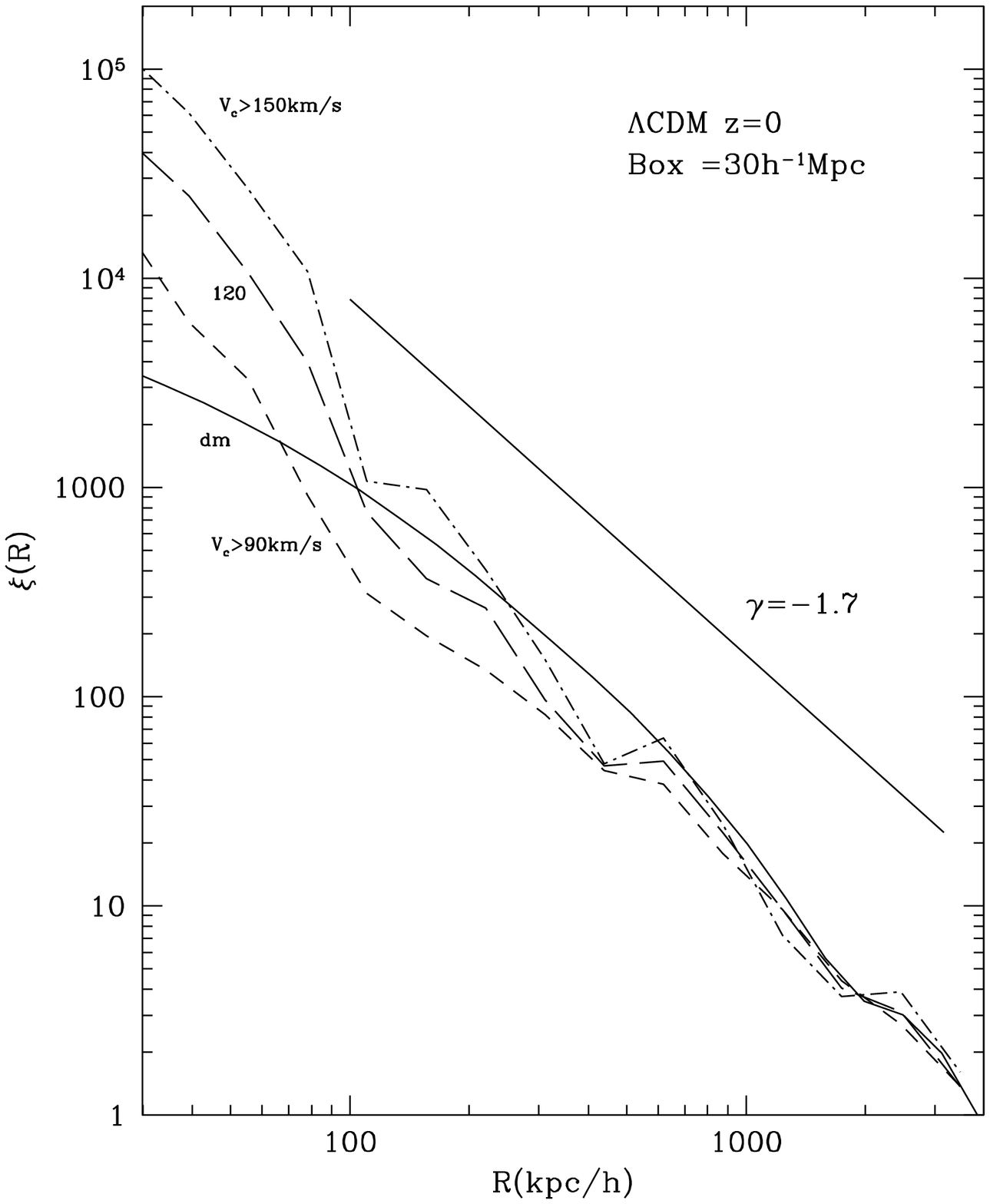}}
\rput[tl]{0}(0.5,1.25){
\begin{minipage}{8.7cm}
  \small\parindent=3.5mm {\sc Fig.}~18.--- Dependence of the galaxy
  correlation function on mass in the $\Lambda$CDM simulation of
  $30\Mpch$ box.  The correlation function increases with the
  rotational velocity, but all curves show the same tendency: positive
  bias on small scales, slight antibias on $(100-1000)\kpch$ scales,
  and no bias on larger scales. Absolute values of the correlation
  functions are affected by the finite box size.
\end{minipage}
}
\endpspicture}
This makes simulations
focused on the individual clusters of the type presented in Ghigna et
al. (1998) a viable alternative.

Unfortunately, even in the case of sufficient resolution, when halos do
survive, the identification of the DM halos in the cores of galaxy
groups and clusters in purely dissipationless simulations remains a
challenge. In the environments that dense, most of halo's dark matter
will be be tidally stripped, which makes it difficult to identify the
leftover on the very dense, smooth background of high-velocity dark
matter particles streaming around and through the halo.  We have
presented two new halo finding algorithms designed to identify
satellite halos located inside the virial radius of a more massive host
halo: the hierarchical friends-of-friends and bound density maxima
algorithms.  Both of our algorithms find practically the same halos,
which are stable (existed at previous moments) and gravitationally
bound.

To exploit the fact that overmerging is (at least to a certain degree)
``beaten'' in our simulation, we consider several statistics of
galaxy-size halos in our simulations and compare them to the
corresponding observed statistics of galaxies.  We use a simple scheme,
based on the empirical Tully-Fisher relation, to assign a luminosity to
the DM halos.  The luminosity function of ``galaxies'' (i.e.,
galaxy-size halos assigned a luminosity) in the $\Lambda$CDM model
reproduces the luminosity function of the CfA catalog (Marzke et al.
1994) reasonably well. Both the simulations and the CfA catalog have an
upturn in the number of faint galaxies ($m_B>-17$). However, magnitudes
of faint ``galaxies'' in the simulations rely on a highly uncertain
extrapolation of the Tully-Fisher relation and on uncertain assumption
about the fraction of elliptical galaxies at these magnitudes. The
number of ``galaxies'' predicted by the CHDM simulation is
significantly higher than in the case of the $\Lambda$CDM simulation
with the same initial random numbers. We failed to produce as nice a
fit to the observational data as for the $\Lambda$CDM simulation.  At
this stage it is difficult to judge if this is a significant problem
for the model or not. Due to the small size of our simulation boxes,
one may argue that simulations with a large box will tend to produce
lower luminosity function keeping at the same time the M/L of galaxy
groups intact. Larger simulations are needed to clarify the situation.

The mass-to-light ratios of galaxy groups in the simulations $\sim
(200-400)h^{-1}$ match those observed reasonably well.  It was
argued (e.g., Bahcall, Lubin, \& Dorman 1995) that dynamics of galaxy
groups favors the low-$\Omega$ Universe. Our results show that
mass-to-light ratios of groups of mass $\lesssim 3\times
10^{13}h^{-1}{\ }{\rm M_{\odot}}$ is insensitive to the matter density.
The halos in the CHDM model are clustered more strongly than the dark
matter and one cannot save the argument for a low-$\Omega$ Universe
by assuming that groups in the CHDM model have too large fraction of
galaxies. It seems that groups occupy too small fraction of the volume
and thus their M/L ratios are not representative for the Universe as a
whole.

Comparison of the halo and matter correlation functions 
indicates that halos are antibiased on 100~kpc -- 1~Mpc
scales. The antibias of the magnitude 0.7--0.9 found in the simulations
is needed for the $\Lambda$CDM model to be compatible with observational
data on the power spectrum in the range of wavenumbers
$k=(0.1-1)h\Mpc^{-1}$ (Klypin et al.  1996; Smith et al. 1998).  We
attribute the antibias to the dynamical friction in groups of
galaxies. The friction tends to drag some galaxies to the very central
part of groups where they merge the central galaxy and disappear (see
Kravtsov \& Klypin 1999 for a more detailed analysis). 

Results of this paper can be used in design of the future numerical
simulations. We have shown that efficient halo finding algorithms can
be developed to identify gravitationally bound satellite halos inside
the virial radius of the other halos. Our analytical estimates and
numerical experiments show that the numerical resolution required to
overcome the overmerging, although quite high, can be achieved with
current numerical codes and computer hardware. The main challenge is
thus purely computational.  This is also true for the simulations that
include dissipative hydrodynamics; while alleviating or obliterating
some of the problems of dissipationless simulations, they are
computationally more intensive. Both numerical approaches have a number
of caveats and potential biases, which could only be avoided with
inclusion of more realistic physics. The latter appears to be
unavoidable, because we cannot reliably predict observed galaxy
properties (and hence mimick the selection criteria of the
observational catalogs) without realistic physics. Fast increase in 
computational capability of modern computers and recent developments of
new efficient numerical algorithms make the perspective for 
advances in this direction look good.

\acknowledgements We thank Joel Primack for careful reading of the
manuscript and comments, Simon White for discussions, and anonymous
referee for constructive criticism and useful suggestions which have
helped to improve content and presentation of the paper. We are
grateful to Avishai Dekel for providing us with computer resources at
the Hebrew University. This work was funded by the NSF and NASA grants
to NMSU, and the collaborative NATO grant CRG 972148. SG acknowledges
support from Deutsche Akademie der Naturforscher Leopoldina with means
of the Bundesministerium f\"ur Bildung und Forschung grant LPD 1996.
Our simulations were done at the National Center for Supercomputing
Applications (Urbana-Champaign, Illinois) and on a Power Challenge
supercomputer at Hebrew University.



\end{document}